\documentclass[12pt,a4paper]{article}            
 \usepackage[skins,theorems]{tcolorbox}
\tcbset{highlight math style={enhanced,
  colframe=red,colback=white,arc=0pt,boxrule=1pt}}
  \usepackage[bookmarksopen, bookmarksnumbered, bookmarksopenlevel=2]{hyperref}
  \usepackage{tikz}
  \usepackage{tikz-3dplot}
  \usepackage{amsthm}
 \usetikzlibrary{calc}
 \usetikzlibrary{decorations} %
 \usepackage[english]{babel}
 \usepackage[toc,page]{appendix}
 \usepackage{amsmath}
 \usepackage{amssymb}
 \usepackage{graphicx}
 \usepackage{hhline}
 \usepackage[bf]{caption}
\usepackage{cite}
\usepackage[vcentermath]{youngtab}
\usepackage{geometry}
\usepackage{slashed}
\usepackage{color}
\usepackage{stackrel}
\usepackage{tikz-cd} 
\usepackage{mathtools}
\usepackage{cancel} 
\usepackage{multirow}
\usepackage[all]{nowidow}

\usepackage{empheq}
\usepackage{arydshln}

\newcommand{\bqa}{\begin{eqnarray}}
\newcommand{\eqa}{\end{eqnarray}}

\newenvironment{eqn*}{\begin{equation*}\begin{aligned}}{\end{aligned}\end{equation*}\noindent}
\hypersetup{
    pdftitle={},
    pdfauthor={},
    pdfsubject={}
}
\numberwithin{equation}{section}
\numberwithin{table}{section}

\makeatletter

\newtheorem*{tadpoleconjecture*}{Tadpole Conjecture}

\newcommand{\be}{\begin{equation}}
\newcommand{\ee}{\end{equation}}
\newcommand{\beq}{\begin{equation}}
\newcommand{\eeq}{\end{equation}}
\newcommand{\ba}{\begin{aligned}}
\newcommand{\ea}{\end{aligned}}

\newcommand{\bea}{\begin{eqnarray}}
\newcommand{\eea}{\end{eqnarray}}

\newcommand{\cN}{\mathcal{N}}

\newcommand{\cM}{\mathcal M}

\newcommand\bi{\begin{itemize}}
\newcommand\ei{\end{itemize}}

\def\unit{{1\kern-.65ex {\rm l}}}
\def\1{{1\kern-.65ex {\rm l}}}

\newcount\hour \newcount\minute
\hour=\time \divide \hour by 60
\minute=\time
\def\now{%
\ifnum \hour<13
  \ifnum \hour=0 \advance \hour by 12 \number\hour:\else \number\hour:\fi%
     \ifnum \minute<10 0\fi%
     \number\minute%
\ A.M.%
\else \advance \hour by -12 \number\hour:%
  \ifnum \minute<10 0\fi%
  \number\minute%
  \ P.M.%
\fi%
}

\makeatother

\begin{document}

\begin{titlepage}
\begin{center}
\rightline{\small }

\vskip 15 mm

{\large \bf
The Tadpole Conjecture in the Interior of Moduli Space
} 
\vskip 11 mm

Severin L\"ust,$^{1}$ Max Wiesner$^{1,2}$

\vskip 11 mm
\small ${}^{1}$ 
{\it Jefferson Physical Laboratory, Harvard University, Cambridge, MA 02138, USA}  \\[3 mm]
\small ${}^{2}$ 
{\it Center of Mathematical Sciences and Applications, Harvard University,\\ Cambridge, MA 02138, USA}

\end{center}
\vskip 17mm

\begin{abstract}
We revisit moduli stabilization on Calabi-Yau manifolds with a discrete symmetry.
Invariant fluxes allow for a truncation  to a symmetric locus in complex structure moduli space and hence drastically reduce the moduli stabilization problem in its dimensionality.
This makes them an ideal testing ground for the tadpole conjecture. For a large class of fourfolds, we show that an invariant flux with non-zero on-shell superpotential on the symmetric locus necessarily stabilizes at least 60\% of the complex structure moduli. In case this invariant flux induces a relatively small tadpole, it is thus possible to bypass the bound predicted by the tadpole conjecture at these special loci. As an example, we discuss a Calabi-Yau hypersurface with $h^{3,1}=3878$
and show that we can stabilize at least 4932 real moduli with a flux that induces M2-charge $N_\text{flux} =3$.
\end{abstract}

\vfill
November 2022
\end{titlepage}

\newpage

\tableofcontents

\setcounter{page}{1}
\section{Introduction}

String Theory famously allows for an enormously large number of possible compactifications, giving rise to a huge Landscape of vacua and effective lower-dimensional models.
The Swampland Program \cite{Ooguri:2006in,Brennan:2017rbf,Palti:2019pca,vanBeest:2021lhn,Grana:2021zvf} aims at finding general criteria that distinguish effective field theories (EFTs) that can be found within the String Theory Landscape from those which fundamentally cannot arise from String Theory.
More generally, it tries to determine which a priori consistent EFTs allow for a UV completion in quantum gravity.%
\footnote{One might of course be tempted to speculate whether these two questions are actually the same.}

Evidence for these criteria (often dubbed ``Swampland Conjectures'') typically stems from two different ways of reasoning:
Either from the bottom-up by studying general properties of gravitational systems, usually the semi-classical physics of black holes, or by collecting and generalizing patterns found in explicit top-down String Theory examples.
The latter approach, however, is often limited by the fact that controlled and calculable examples are mostly only available at weak coupling and/or unbroken supersymmetry.
This of course poses the risk of extrapolating from specific properties of weakly-coupled, supersymmetric compactifications to the whole landscape, falling victim to a ``lamppost effect''.
Despite these dangers, weak coupling limits play a particularly prominent role within the Swampland Program.
It is generally believed that quantum gravity theories do not allow for any free parameters.
Therefore, coupling constants must be dynamical and realized as the expectation values of scalar fields.
This implies that every weak coupling limit corresponds to an asymptotic boundary of the scalar field space.
In particular, the points where a coupling constant vanishes and a global symmetry would be restored lie at infinite distance from any point in the interior of the scalar field space \cite{Ooguri:2006in}.

Even though conceptually important, the often high-dimensional moduli spaces of Calabi--Yau compactifications pose a serious challenge for string phenomenology.
Large numbers of massless (or light) scalar fields are not observed and must be stabilized if one wants to obtain stable, realistic string theory vacua.
For the complex structure moduli of IIB or F-theory compactifications this can be achieved by introducing non-trivial three or four-form fluxes on the Calabi--Yau manifold's middle cohomology \cite{Grana:2005jc,Douglas:2003um}.
The large combinatorical number of such fluxes on topologically complicated Calabi--Yau manifolds has led to statistical estimates on the enormous size of the String Theory Landscape \cite{Denef:2004ze,Denef:2004cf}. However, the intuition that any choice of fluxes, even if supported on only a small number of cycles, can give rise to a consistent vacuum where a large number of moduli is stabilized has recently been challenged by the ``tadpole conjecture'' \cite{Bena:2020xrh}.
This conjecture states that the fluxes that are needed to stabilize a large number of moduli carry a charge that is directly proportional to the number of stabilized moduli.
As this charge contributes to the tadpole cancellation condition and is therefore bounded from above, this statement -- if true -- drastically reduces the size of the landscape of viable vacua.

For phenomenological purposes and in order to obtain a stable compactification, it is is generally desirable to stabilize all complex structure moduli by fluxes, with no (or at most a compact) residual moduli space of unstablized moduli.
On the other hand, one might ask whether the linear scaling behavior of the tadpole conjecture persists also for partial moduli stabilization.
In this work, we therefore suggest to refine the tadpole conjecture into a stronger and a weaker version. The stronger version states that for the stabilization of any number, $n_{\rm stab}$, of moduli the required flux induces an M2-brane tadpole is proportional to $n_{\rm stab}$. Instead, the weaker condition only requires that in case \emph{all} complex structure moduli of a given fourfold are to be stabilized, the required flux has to induce a tadpole of the order of $h^{3,1}$. 
Clearly, the weaker version is implied by the stronger one.

First evidence for the tadpole conjecture was provided by an extensive, computer-aided search on K3$\times$K3 backgrounds \cite{Bena:2021wyr} where it was found that full moduli stabilization can only be obtained within the tadpole bound at special, singular gauge-enhancement points deep in the interior of moduli space.
Yet soon after, general stabilization schemes were found that seemingly violate the tadpole conjecture in the large complex structure regime of general three and four-folds \cite{Marchesano:2021gyv}.
However, it was pointed out in \cite{Plauschinn:2021hkp,Lust:2021xds,Grimm:2021ckh} that in order to retain control over the large complex structure expansion generically fluxes with large charges are needed, compatible with the tadpole conjecture.
Notice, that the requirement of control over the large complex structure expansion is the same as staying sufficiently close to the asymptotic boundary of scalar field space.
Subsequently, using recent insights on asymptotic Hodge structure, the linear scaling relation between the induced charge of the fluxes and the number of stabilized moduli was confirmed for all strict asymptotic limits in complex structure moduli space \cite{Grana:2022dfw}.

Consequently, and in the spirit of the discussion above, the question naturally arises if the tadpole conjecture remains generally true even in the interior of moduli space or if it is only a consequence of stabilizing moduli in its asymptotic regions.
This is the motivating question of this paper.
Unfortunately, explicitly stabilizing a large number of moduli in the interior of the scalar field space, far away from any asymptotic regime such as the large complex structure region is a difficult task.
The required period integrals are typically only known for Calabi--Yau manifolds with a small number of moduli or in the large complex structure limit and are generally expensive to compute.

Therefore, we rely on a technique that has already been used frequently in earlier examples of explicit flux compactifications \cite{Giryavets:2003vd,Denef:2004dm,Cicoli:2013cha,Demirtas:2019sip, Blanco-Pillado:2020wjn,Blanco-Pillado:2020hbw, Braun:2020jrx}.
Calabi--Yau manifolds that exhibit a discrete symmetry acting on their moduli spaces allow for a consistent truncation on the invariant locus in moduli space.
This requires a choice of fluxes that are invariant under the discrete symmetry group and reduces the problem drastically in its dimensionality.
Instead of solving the vacuum (F-term) equations for all moduli explicitly, one only needs to solve them for the relatively small subset of invariant moduli.
The vacuum equations of the non-invariant moduli are solved automatically by virtue of the discrete symmetry. Still, a solution to the vacuum equations does not guarantee stabilization of the non-invariant moduli since there may remain directions orthogonal to the invariant locus that are not obstructed by the flux-induced potential.
For special points along the invariant locus, namely the Fermat point, the number of directions obstructed by invariant flux have been counted in \cite{Braun:2020jrx,Becker:2022hse} with the result that the scaling between number of stabilized moduli and the tadpole is indeed as required by the tadpole conjecture.

In this work, we address the tadpole conjecture at generic points along the invariant locus, away from the Fermat point. To that end, we do not directly address the moduli stabilization problem in type IIB/F-theory but focus on Calabi--Yau fourfold compactifications of M-theory with $G_4$-flux. We further concentrate entirely on the complex structure sector by restricting to primitive $G_4$-fluxes. Thus, if the fourfold is elliptically fibered, we can translate our results directly to 4d F-theory compactification on the same fourfold. To exploit the symmetry arguments outlined above, we focus on a special class of fourfolds obtained as hypersurfaces in weighted projective space. There are 3462 of such Calabi--Yau fourfolds which typically have a very high-dimensional complex structure moduli space $h^{3,1}\sim \mathcal{O}(1000)$ making them ideally suited to discuss stabilization of large number of complex structure moduli. The loci in complex structure moduli space along which there is an enhanced discrete symmetry, on the other hand, have relatively small dimension of $h^{3,1}_{\rm inv.}\sim \mathcal{O}(10)$. Therefore the reduced moduli stabilization problem on the invariant locus by means of invariant fluxes becomes a tractable problem. 

The question pertinent to our analysis in this paper is whether a choice of $G$-invariant flux, that leads to a solution to the F-term equation with $W_0\equiv W_{D_a W=0}\neq 0$, is capable of stabilizing a large numer of non-invariant moduli. To that end, we do not assume the vacuum to be located at any particular point along the invariant locus. In this regard our analysis differs from previous work \cite{Braun:2020jrx,Becker:2022hse}. Of particular interest for us is the effect of the non-zero $W_0$ for the stabilization of the non-invariant complex structure moduli. As was noticed already in \cite{deAlwis:2013jaa}, a non-zero $W_0$ automatically obstructs half of the real deformations such that the residual moduli space has at most real dimension $h^{3,1}$. We therefore propose that in this case the tadpole conejecture should relate the flux-induced tadpole to the number, $n_{\rm stab}$, of real moduli that are not automatically stabilized. Still, naively a non-zero $W_0$ yields an obstruction to all deformation away from the invariant locus. However, additional obstructing terms can originate from the Hessian $D^2W$ which in principle can cancel an obstruction induced by $W_0\neq 0$ if the rank of $D^2 W$ is sufficiently large.

To see whether it is conceivable that a full cancellation takes place we perform a scan over (a certain subset) of the hypersurface Calabi--Yau fourfolds in weighted projective space and numerically compute the rank of $D^2 W$ at a generic point along the invariant locus. As a result we provide a lower bound on the number $n_{\rm stab}$ of non-invariant directions that are obstructed by a choice of a self-dual invariant flux leading to $W_0\neq 0$: 
\begin{equation}
    n_{\rm stab} \gtrsim 0.2 \,(h^{3,1} -h^{3,1}_{\rm inv.})\,.
\end{equation}
 Thus, for $h^{3,1}\gg h^{3,1}_{\rm inv}$ it is possible to stabilize a large number of complex structure fields just by solving the reduced stabilization problem along the invariant locus. We illustrate that this is indeed possible in an explicit example where it is possible to stabilize at least $n_{\rm stab}=1052$ additional real moduli with a flux leading to a tadpole of $N_{\rm flux}=3$.
 This shows explicitly that it is indeed possible to avoid the bound imposed by the strong form of the tadpole conjecture at special loci in the interior of the moduli space. 

The rest of this paper is organized as follows:  In section~\ref{sec:review} we review M-theory flux compactifications and the tadpole conjecture. In section~\ref{sec:fermatCY} we provide some basics of Calabi--Yau hypersurfaces in weighted projective space and their complex structure moduli spaces. In section~\ref{sec:symmetrictadpoleconjecture} we challenge the tadpole conjecture by studying the stabilization of complex structure deformations through symmetric fluxes. To illustrate the difference between the asymptotic regions and the interior points in the moduli space, we compare the moduli stabilization at the symmetric locus with the situation in asymptotic regimes in section~\ref{sec:comparison}. Finally, we discuss our results in section~\ref{sec:conclusions}. 

\section{Flux vacua and the tadpole problem}\label{sec:review}
In this work we are interested in flux compactifications of M-theory/F-theory on Calabi--Yau fourfolds $Y_4$. The relevant flux in this case is the M-theory four-form flux, $G_4$, which has to be quantized such that \cite{Witten:1996md}
\begin{equation}\label{eq:quantization}
    G_4 +\frac12 c_2(Y_4) \in H^4(Y_4,\mathbb{Z}). 
\end{equation}
The middle cohomology for fourfolds can in general be split into a vertical part, a horizontal part and a remainder that is neither vertical nor horizontal \cite{Braun:2014xka}. The horizontal part can be generated by varying the holomorphic four-form $\Omega$ and allows for an orthogonal decomposition as 
\begin{equation}
    H^4_{\rm hor.} = H^{4,0} \oplus H^{3,1} \oplus H^{2,2}_{\rm hor.} \oplus H^{1,3} \oplus H^{0,4}\,. 
\end{equation}
Of particular interest to us are primitive fluxes satisfying $G_4\wedge J=0$ for $J$ the K\"ahler form on $Y_4$.\footnote{In case the second Chern class of $Y_4$ is not even, the quantization condtion \eqref{eq:quantization} forces us to add a vertical component to the flux. In the following we ignore any such component and only focus on the primitive component of $G_4$ which by \eqref{eq:quantization} has to be integer quantized.} For a primitive $G_4$-flux, the induced classical superpotential is given by \cite{Gukov:1999ya}
\begin{equation}
 W = \int_{Y_4} G_4 \wedge \Omega \,, 
\end{equation}
which through $\Omega$ depends on the complex structure deformations $\Phi^\alpha$ of $Y_4$ of which there are $h^{3,1}$. In this work, we exclusively focus on the complex structure sector and ignore any mixing with the K\"ahler sector. The F-term supersymmetry conditions 
\begin{equation}
D_\alpha W = 0 \,,
\end{equation}
for the complex structure moduli are solved by self-dual four-form flux  $G_4=*G_4$.\footnote{If we want to solve \emph{all} F-term equations including the K\"ahler direction and hence get actual supersymmetric vacua, the condition on the fluxes is more constraining than just self-duality as has been discussed in \cite{Lust:2022lfc}.} In addition, for a consistent compactification the M2-brane tadpole needs to be cancelled. For M-theory on four-folds the tadpole cancellation condition reads 
\begin{equation}\label{tadpole}
    \frac12 \int_{Y_4} G_4\wedge G_4 +N_{\rm M2} = \frac{1}{24}\chi(Y_4)\,,
\end{equation}
where $N_{\rm M2}$ is the number of space-time filling M2-branes and $\chi(Y_4)$ denotes the Euler-characteristic of $Y_4$. Since for a supersymmetric theory $N_{\rm M2}\geq 0$, the possible flux choices are constrained by 
\begin{equation}\label{tadpolebound}
    N_\text{flux} \equiv \frac12 \int_{Y_4}G_4\wedge G_4 \leq\frac{\chi(Y_4)}{24}\,. 
\end{equation}
Of particular interest for us is the dimension of the locus $\mathcal{M}_{\rm vac}$ in field space along which $D_\alpha W=0$ does have a solution for a given $G_4$ flux. More precisely, we consider the number of directions in moduli space that are obstructed at first order, i.e., that correspond to massive directions. In the following we refer to this number as the \emph{co-dimension} of $\mathcal{M}_{\rm vac}$.
In \cite{Becker:2022hse} this 
statement was formulated in a mathematically more precise  way and expressed in terms of the Zariski dimension of $\mathcal{M}_{\rm vac}$. 

In order to find this co-dimension let us assume that we found a solution to the equations $D_aW =0$ at a point $\Phi_{\rm crit}$ in moduli space. Then the number of first-order obstructed directions, i.e.\ the co-dimension of $\mathcal{M}_{\rm vac}$, is determined by the rank of the matrix 
\begin{equation}\label{defM}
  M = \left.\left(\begin{matrix} D_\alpha D_\beta W & \bar{D}_{\bar \alpha} D_\beta W \\ D_\alpha \bar{D}_{\bar \beta} \bar W & \bar{D}_{\bar \alpha} \bar{D}_{\bar \beta} \bar W \end{matrix}\right)\right|_{\Phi_{\rm crit}}=\begin{pmatrix}
      D_\alpha D_\beta W & g_{\bar \alpha  \beta} W \\ g_{ \alpha \bar \beta}\bar{W}&\bar{D}_{\bar \alpha} \bar{D}_{\bar \beta} \bar W 
  \end{pmatrix}\bigg|_{\Phi_{\rm crit}}\,,
\end{equation}
where in the last step we used holomorphicity of $W$.
The discussion of the rank of $M$ depends significantly on whether $W_0$, i.e.\ the value of $W$ at $\Phi_{\rm crit}$, vanishes or not. If $W_0=0$, the rank of the matrix $M$ is only determined by the rank of $D_\alpha D_\beta W$ and the residual moduli space is complex. On the other hand, as pointed out in \cite{deAlwis:2013jaa}, if $W_0\neq 0$ and the metric $g_{\alpha \bar{\beta}}$ is non-degenerate, the residual moduli space is real and at most of dimension $h^{3,1}$. 
However, the actual dimension of $\cM_{\rm vac}$ again depends on the details of $D_\alpha D_\beta W$. 

In general, one would expect that 
the flux $G_4$ needs to be sufficiently complicated 
in order to solve the F-term conditions $D_\alpha W = 0$ while at the same time ensuring that $D_\alpha W$ is non-constant in many directions. However, such a complicated flux would typically lead to a large contribution to the M2-brane tadpole and
stabilizing many moduli
may thus not be possible within the bound \eqref{tadpolebound}. In \cite{Bena:2020xrh} it has been conjectured that the co-dimension of $\mathcal{M}_{\rm vac}$, or equivalently the number, $n_{\rm stab}$, of stabilized moduli, and the flux-induced tadpole are related linearly. In its original formulation \cite{Bena:2020xrh} considers $n_{\rm stab}$ to be the number of stabilized \emph{complex} fields. However, given our previous discussion, for $W_0\neq 0$ it is more natural to count real scalar fields. In particular, half of the real directions in moduli space are automatically stabilized. Thus, in order to formulate the tadpole conjecture in a sensible way, in the case $W_0\neq 0$ we should only count the real directions that are stabilized \emph{in addition} to those stabilized automatically. In the following, we will therefore work with the following form of the tadpole conjecture: 

\begin{tadpoleconjecture*}[Strong Form]\!\!\footnote{A related formulation of the tadpole conjecture for partial moduli stabilization (without distinguishing between the cases $W_0=0$ and $W_0 \neq 0$) was recently put forward in \cite{Becker:2022hse}.}
The fluxes that stabilize a large number, $n_{\rm stab}\gg1$, of complex structure moduli in a smooth Calabi--Yau fourfold compactification of M-theory lead to an M2-brane tadpole that grows at least linearly in $n_{\rm stab}$, i.e. 
\begin{equation}\label{tadpoleconj}
    N_{\rm flux} \gtrsim \alpha \,n_{\rm stab}\,,\qquad \alpha\sim \mathcal{O}(1)\,,
\end{equation}
where $n_{\rm stab}$ is defined as 
\begin{subequations}
    \begin{align}
       \label{nstabMink} W_0=0:\qquad &n_{\rm stab} \equiv \operatorname{codim}_{\mathbb{C}}\left(\cM_{\rm vac}\right)\,,\\
        \label{nstabAdS} W_0\neq 0:\qquad &n_{\rm stab}\equiv \operatorname{codim}_{\rm \mathbb{R}}\left(\cM_{\rm vac}\right) - h^{3,1}\,.
    \end{align}
\end{subequations}
\end{tadpoleconjecture*}

\noindent In both cases $n_{\rm stab}$ can take values in $\{0, \dots, h^{3,1}\}$ with full moduli stabilization corresponding to $n_{\rm stab}=h^{3,1}$. In fact, oftentimes the tadpole conjecture is interpreted as an obstruction to stabilize \emph{all} complex structure moduli of a given Calabi--Yau fourfold. This motivates a formulation of a weaker form of the tadpole conjecture that only applies to the case $n_{\rm stab}=h^{3,1}$:

\begin{tadpoleconjecture*}[Weak Form] For a fixed, smooth Calabi--Yau fourfold with $h^{3,1}\gg 1$, the fluxes that stabilize \emph{all} complex structure moduli (i.e.~$n_\mathrm{stab}=h^{3,1}$) induce an M2-brane tadpole that is bounded as 
\begin{equation}\label{tadpoleconj2}
    N_{\rm flux}\gtrsim \beta\,h^{3,1} \,,\qquad \beta\sim \mathcal{O}(1)\,. 
\end{equation}
\end{tadpoleconjecture*}

\noindent
The tadpole conjecture has been investigated in various settings starting with \cite{Bena:2021wyr}. Much focus has been on confirming or disproving the stronger version of the tadpole conjecture in asymptotic limits in field space \cite{Marchesano:2021gyv, Plauschinn:2021hkp, Lust:2021xds,Grimm:2021ckh, Grana:2022dfw}. Instead, in the following we want to study the validity of the two forms of the tadpole conjecture away from asymptotic limits in the interior of the moduli space. Progress in this direction has recently been made in the type IIB context for orientifold compactifications with $h^{1,1}=0$ \cite{Becker:2022hse}. Here, we focus on a class of Calabi--Yau fourfolds that generically has $h^{3,1}\gg 1$ but still allows for explicit moduli stabilization at special points in the interior of the moduli space. We introduce some basic properties of these fourfolds in the next section before we study moduli stabilization in section \ref{sec:symmetrictadpoleconjecture}. 
\section{Hypersurface Calabi--Yau fourfolds}\label{sec:fermatCY}

In this and the following section, we  focus on a simple class of Calabi--Yau fourfolds that can be obtained as a hypersurface in a weighted projective space 
\begin{equation}\label{projspace}
    \mathbb{P}^5_{a_1,\dots, a_6} =  \left(\mathbb{C}^6\backslash \{0\}\right)/\,\mathbb{C}^\star\,,
\end{equation}
where the $\mathbb{C}^\star$ acts on the coordinates $x_{n=1,\dots, 6}$ of $\mathbb{C}^6$ as
\begin{equation}\label{eq:lambdascaling}
  (x_1,\dots, x_6) \sim (\lambda^{a_1} x_1,\dots , \lambda^{a_6}x_6)\,.
\end{equation}
A hypersurface in this weighted projective space can be described as the zero locus of a polynomial $P$ for which the Calabi--Yau condition reads 
\begin{equation}\label{degreeP}
 d\equiv \sum_{n=1}^6 a_n =\text{deg}\, P\,, 
\end{equation}
where $\text{deg}\, P$ is the degree of $P$ under the re-scaling \eqref{eq:lambdascaling}. A simple class of Calabi--Yau fourfolds $Y_4$ can hence be obtained from Fermat-type polynomials of the form 
\begin{equation}
 P_0 = \sum_{n=1}^6 x_n^{d/a_n}\,.
\end{equation}
For the Calabi--Yau defined by $P_0=0$ to be transverse, i.e.~$P_0=0$ and $dP_0=0$ not to have a common solution, we have to require that each $a_n$ is a divisor of $d$ \cite{Candelas:1989hd}. Based on these Fermat-type Calabi--Yau four-folds we can further consider the families of CYs obtained by deformations of $P_0$ of the form 
\begin{equation}\label{P0deform}
    P= P_0 + \sum_{\underline{\alpha}} \Phi_{\underline{\alpha}} \prod_{n=1}^6 x_n^{\alpha_n} \,,
\end{equation}
where we sum over $\underline{\alpha}=(\alpha_1,\dots, \alpha_6)$ subject to the constraint $$\sum\limits_{n=1}^6\alpha_n a_n = d\,.$$ The complex parameters $\Phi_{\underline{\alpha}}$ can be interpreted as complex structure deformations of the Calabi--Yau defined by $P_0$. Notice that not all $\underline{\alpha}$ lead to independent deformations of $P_0$. To single-out the independent deformations, we need to divide out the ideal of $P_0$ generated by 
\begin{equation}
    \langle \partial_{x_n} P_0 \rangle = \langle x_1^{d/a_1-1},\dots , x_6^{d/a_6-1} \rangle\,. 
\end{equation}
We denote the number of independent polynomial deformations of $P_0$ by $h^{3,1}_{\rm poly}$. In general, the $\mathbb{C}^*$-action of the weighted projective space in \eqref{projspace} has fixed points on the hypersurface $\{P=0\}$ leading to orbifold singularities. These singularities can be resolved by turning on vevs for massless twisted states leading eventually to a smooth CY which has additional moduli associated to these resolution modes.\footnote{Whereas for threefolds all resolution modes have a geometric interpretation \cite{Greene:1991we} this is not necessarily the case for fourfolds \cite{Klemm:1996ts}.} The total number of deformations can be obtained from the Landau--Ginzburg formula \cite{Vafa:1989ih,Vafa:1989xc}
\begin{equation}\label{eq:trace}
\text{tr}\; t^{d J_0} \bar{t}^{d\bar J_0} = \sum_{l=0}^{d-1}\prod_{l\frac{a_n}{d}\in \mathbb{Z}}\frac{1-(t\bar{t})^{d-a_n}}{1-(t\bar t)^{a_n}} \prod_{l\frac{a_n}{d}\notin \mathbb{Z}} (t\bar t)^{d/2-a_n}\left(\frac{t}{\bar t}\right)^{d\left(l\frac{a_n}{d} \;\text{mod}\; \mathbb{Z}-1/2\right)}\,.
\end{equation}
Here $t$ is an auxiliary variable, $d$ the degree of $P$ as in \eqref{degreeP} and $a_n$ are the weights of the weighted projective space as in \eqref{eq:lambdascaling}. The degeneracy of states with $(J_0, \bar{J}_0)$ charges $(d-p,q)$ gives the Hodge number $h^{p,q}$ of the resolved CY fourfold. Depending on the details of the resolution, the twisted moduli can contribute to both $h^{3,1}$ or $h^{1,1}$. 

In this note, we will merely focus on the polynomial deformations introduced in \eqref{P0deform}. Among the families of CYs given by the hypersurface $\{P=0\}$, the Fermat-type CYs are special since they are invariant under a discrete symmetry group $G$. We can write the generators of $G$ as $g^{(k)}=\left(g_1^{(k)}, \dots, g_6^{(k)}\right)$ which act on the coordinate $x_n$ of $\mathbb{P}^5_{a_1,\dots, a_6}$ as 
\begin{equation}\label{gaction}
    g^{(k)}:\; x_n \mapsto \exp\left(2\pi i g_n^{(k)} \frac{a_n}{d} \right)x_n\,.
\end{equation}
For the Fermat CY to be symmetric under the action of $g^{(k)}$ the $g^{(k)}_n$ further need to satisfy
\begin{equation}\label{generatorcond}
 \sum_{n=1}^6 g_n^{(k)} a_n =0 \,.
\end{equation}
Since the action of the generators $g^{(k)}$ has to be understood modulo the projective rescaling $x_n\rightarrow \lambda^{a_n} x_n$ not all $g^{(k)}$ satisfying \eqref{generatorcond} are in fact independent. In section~\ref{sec:ex} we illustrate this for an example with $(a_1,\dots,a_6)=(1,1,1,1,8,12)$.

The deformations of the polynomial $P_0$ as in \eqref{P0deform} are in general not $G$-invariant unless 
\begin{equation}
 \sum_{n=1}^6 \alpha_n g_n^{(k)} \frac{a_n}{d} \in \mathbb{Z}\,. 
\end{equation}
Let us denote by $h^{3,1}_{\rm inv}$ the number of such $G$-invariant polynomials. We may then split the complex scalar fields $\Phi_{\underline{\alpha}}$ into two sets 
\begin{equation}
 \Phi_{\underline{\alpha}} \rightarrow (\psi^a, \phi^i)\,,
\end{equation}
where the $\psi^{a}$ parametrize the $G$-invariant deformations whereas $\phi^{i}$ are the parameters associated to polynomial deformations of $P_0$ not invariant under $G$. We are particularly interested in the locus within the deformation space of $P_0$ defined by $\phi^i=0$. The polynomial $P$ in \eqref{P0deform} is $G$-invariant along this locus to which we refer to as \emph{symmetric locus}. Recall that in addition to the polynomial deformations of $P$ there can be further complex structure deformations associated to the resolution of orbifold singularities. In general the group $G$ does not have a well-defined action on these complex structure moduli. Since in the following we want to use the symmetry $G$ to constrain the periods the Calabi--Yau fourfold we restrict to the case that the resolution of the hypersurface $P=0$ does not yield any additional complex structure deformations. We hence restrict to the case $h^{3,1}=h^{3,1}_{\rm poly}$.

\subsubsection*{The symmetric locus in moduli space}

Given a CY of the type discussed above with $h^{3,1}=h^{3,1}_{\rm poly}$ we now specialize to the vicinity of the symmetric locus $(\psi^a, \phi^i)=(\psi^a,0)$. Under $g\in G$ the non-invariant deformations $\phi^i$ transform as 
\begin{equation}\label{transform}
    \phi^i \,\stackrel{g}{\rightarrow}\,\beta_{i}(g)\, \phi^i\,, \qquad \text{no summation,}
\end{equation}
for some discrete complex phases $\beta_i(g)$. In a similar way to the deformations, we can split the periods of the holomorphic $(4,0)$-form $\Omega$ into invariant and non-invariant periods as\footnote{In principle the action of $G$ on the periods could also be accompanied by a non-trivial K\"ahler transformation as pointed out in \cite{Cicoli:2013cha}. However, in the remainder of this work we only consider K\"ahler-covariant expressions such that our results are independent of this subtlety.} 
\begin{equation}\label{def:periods}
    (\Pi^A,\tilde{\Pi}^I) \stackrel{g}{\rightarrow}\, (\Pi^A,\gamma_{I}(g)\, \tilde{\Pi}^I)\,, \qquad \text{no summation.}
\end{equation}
Here, the $\gamma_{I}(g)$ are again complex phases and the indices $A$ and $I$ run over a basis of invariant and non-invariant four-cycles, respectively. Notice that the basis $(\Pi^A,\tilde{\Pi}^I)$ does not necessarily correspond to an integer basis of periods, but for our purposes is just \emph{some} basis of periods with nice transformation properties under $G$.
For the following discussion, it is, however, important to note that there exists an integer basis that is compatible with this split of the periods.

Given the transformation properties of the deformations in \eqref{transform}, we can infer the dependence of the periods on the $\phi^i$. For instance, around the locus $\phi^i = 0$ we have 
\begin{equation}\label{Aperiods}
    \Pi^A = \pi^{A,0}(\psi^a) + \pi_{ij}^{A,2}(\psi^a) \phi^i \phi^j + \dots \,,
\end{equation}
where $\pi_{ij}^{A,2}(\psi^a)$ is only non-zero if $\beta_i(g) \beta_j(g)=1$ for all $g\in G$. On the other hand, the dependence of $\pi_{ij}^{A,2}$ on the invariant deformations $\psi^a$ cannot be determined purely from symmetry considerations. In principle, it is therefore possible that $\pi^{A,2}_{ij}$ vanishes identically even if $\beta_i(g) \beta_j(g)=1$ for all $g\in G$. On the other hand, the periods $\tilde{\Pi}^I$ have the expansion 
\begin{equation}\label{PiIexpansion}
    \tilde{\Pi}^I = \tilde{\pi}^{I,1}_i(\psi^a) \phi^i +\tilde{\pi}^{I,2}_{ij}(\psi^a)\phi^i \phi^j +\tilde{\pi}^{I,3}_{ijk} (\psi^a) \phi^i \phi^j  \phi^k + \tilde{\pi}^{I,4}_{ijkl} (\psi^a) \phi^i \phi^j  \phi^k\phi^l +\dots\,. 
\end{equation}
At each order in the expansions, the coefficients $\pi^{A,2}_{ij},\tilde{\pi}^{I,2}_{ij}, \tilde{\pi}^{I,3}_{ijk}$, etc.\ are not all independent but satisfy certain linear relations. To see this we notice that these terms can be associated to the squares (cubes) of the polynomials in \eqref{P0deform}. But these polynomials are defined modulo the ideal $\langle \partial P\rangle$. At a given point along the symmetric locus $\{\phi^i=0\}$ the ideal depends on $\psi^a$ dictating the relative $\psi^a$-dependence of the different expansion coefficients in $\Pi^A, \tilde{\Pi}^I$.
Moreover, the expansion coefficients can be arranged into representations of the outer automorphism group of $G$.
Invariance with respect to outer automorphism of $G$ requires all coefficients that lie in the same representation to have the same functional dependence on $\psi^a$, reducing the number of independent components further.

Notice that along the symmetric locus $\{\phi^i=0\}$ all non-invariant periods $\{\tilde{\Pi}^I\}$ vanish as there is no constant term in the expansion \eqref{PiIexpansion} since such a term would spoil the transformation \eqref{def:periods}. However, the vanishing of $\tilde{\Pi}^I$ does not lead to a singularity in field space with additional massless degrees of freedoms as, e.g., there is no supersymmetric cycle that vanishes at the supersymmetric point. To see this, note that $dP\neq 0$ for generic $\psi^a$ and hence the Calabi--Yau fourfold defined by $P=0$ is non-singular along the locus $\{\phi^i=0\}$, indicating that the volume of all 4-cycles is non-vanishing.\footnote{For primitive four-cycles the volume is measured by $e^{K/2}\int |\Omega|$ which can be non-zero even if all period integrals $\int \Omega$ vanish as it happens e.g.\ at the Fermat point.} Alternatively, we can consider the field space metric 
\begin{equation} 
g_{i \bar \jmath}= -\partial_i \partial_{\bar \jmath}\left( \log \int_{Y_4} \Omega \wedge \bar\Omega\right)\,. 
\end{equation}
Given the expansion of the periods in terms of the deformations, we infer that to leading order
\begin{equation}\label{OmegawedgebarOmega}
     \int_{Y_4} \Omega \wedge \bar{\Omega} = K^{(0)}(\psi^a,\bar{\psi}^b) + K^{(2)}_{i\bar\jmath}(\psi^a,\bar\psi^b) \phi^i \bar{\phi}^{\bar \jmath} + \left(K_{ij}^{(2)}(\psi^a,\bar{\psi}^b)\phi^i\phi^j + \text{c.c.}\right)\,.
\end{equation}
The K\"ahler potential is invariant under $G$ up to a K\"ahler transformation. Since $K_{i\bar \jmath}^{(2)}$ is a function of the invariant moduli $\psi^a$ only, it is invariant under $G$. Thus $K_{i\bar \jmath}^{(2)}=0$ unless the product $\phi^i \bar \phi^{\bar \jmath}$ is invariant under $G$. Similarly, $K_{ij}^{(2)}$ can only be made up of terms stemming from the invariant periods $\Pi^A$, i.e. 
\begin{equation}
  K_{ij}^{(2)} (\psi^a, \bar \psi^b) = \pi^{A,2}_{ij} \eta_{AB}\bar{\pi}^{B,0}\,,
\end{equation}
where $\eta_{AB}$ is the intersection form restricted to the invariant sector. From the K\"ahler potential, we can derive the metric on field space along the symmetric locus
\begin{equation}
    g_{i\bar \jmath} = \frac{K^{(2)}_{i\bar\jmath}(\psi^a,\bar\psi^b)}{ K^{(0)}(\psi^a,\bar{\psi}^b)^2}\,,
\end{equation}
which for generic $K^{(2)}_{i \bar \jmath}$ and generic values of $\psi^a$ is non-degenerate and in particular does not feature a logarithmic divergence in the $\phi^i$ as would be expected for a point in field space where a supersymmetric cycle vanishes. 

Setting the non-invariant deformations $\phi^i=0$, we are left with a lower-dimensional deformation space spanned by the $\psi^a$. In certain cases, this space can be identified with the \emph{full} complex structure moduli space of the mirror $X_4$ of $Y_4$. To see this, let us recall that by the Greene--Plesser construction \cite{Greene:1990ud} the mirror of $Y_4$ can be obtained by quotienting the hypersurface $P=0$ by the group $G$, i.e.~$X_4=\{P=0\}/G$. Of course, in order to be able to quotient by $G$, the hypersurface $P=0$ needs to have $G$ as one of its symmetries which is precisely the case along $\{\phi^i=0\}$. The $G$-orbifolding renders the deformations $\phi^i$ massive whereas the deformations $\psi^a$ remain as massless complex structure deformations of $X_4$. Still, when orbifolding by $G$ additional singularities are induced. Resolving these orbifold singularities might again introduce additional complex structure deformations. Therefore in general the complex structure moduli space of the mirror $X_4$ is larger than just the invariant locus spanned by the $\psi^a$. However, in case there are no further complex structure deformations of $X_4$ associated to the resolution of the orbifold singularities the symmetric locus $\{\phi^i=0\}$ of $Y_4$ is identical to the complex structure moduli space of $X_4$. This implies in particular that in this case the invariant periods $\Pi^A$ can be identified with the periods of the mirror $X_4$.

Given an integer basis of periods $\Pi^A$ of $X_4$ one might be worried whether these also correspond to an integer basis of periods on the invariant locus on $Y_4$ or whether an integer basis of periods of $Y_4$ is related to that of $X_4$ via a rescaling by the order $|G|$ of $G$. To see that this is not case, let us take a closer look at the effect of the orbifolding. Therefore consider the holomorphic $(4,0)$-form $\Omega$ which can be chosen to be of the form (cf. \cite{Candelas:1990rm,Candelas:1993dm} for the analogous CY 3-fold case) 
\begin{equation}
    \Omega = -\frac{f(\psi^a)}{(2\pi i)^4} \frac{x_6 dx_1dx_2dx_3 dx_4}{\frac{\partial P}{\partial x_5}}\,,
\end{equation}
for some suitable function $f$ of the invariant deformations $\psi^a$. Using this $(4,0)$-form for both, $Y_4$ and its mirror $X_4$, when integrating $\Omega$ over a basis of cycles in $X_4$ the orientifold identifications will lead to an extra factor of $|G|^{-1}$ \cite{Candelas:1990rm,Candelas:1993dm}. To identify the periods of $Y_4$ with the periods of $X_4$ we should therefore rescale $\Omega(X_4)\rightarrow |G|\, \Omega(Y_4)$. A rescaling of $\Omega$ does, however, not change the physics and in particular not the quantization condition. Therefore an integer basis of periods on $X_4$ indeed leads to an integer basis of periods on $Y_4$ without the need to introduce extra factors of $|G|$.  Notice that since the periods of $Y_4$ on the locus $\{\phi^i=0\}$ are identified with the periods of $X_4$ also $K^{(0)}(\psi^a, \bar\psi^b)$ appearing in \eqref{OmegawedgebarOmega} is just the K\"ahler potential on the mirror $X_4$.

To summarize, along the symmetric locus in the complex structure moduli space of $Y_4$ the moduli field space metric is non-degenerate in the $\phi^i$ directions indicating no additional degrees of freedom. On the other hand, in the special cases discussed above, the remaining directions along the symmetric locus can be identified with the complex structure moduli space of the mirror $X_4$ of $Y_4$. We will use these properties in the next section to analyze flux vacua along this symmetric locus. 

\section{Moduli stabilization at the symmetric locus}\label{sec:symmetrictadpoleconjecture}
We now turn to flux compactifications of M-/F-theory on the Calabi--Yau fourfold $Y_4$ obtained as the hypersurface $P=0$ in the weighted projective space $\mathbb{P}_{a_1,\dots, a_6}^6$. Therefore let us expand $G_4= m_A \alpha^A + \tilde{m}_I \tilde{\alpha}^I$ in a basis of four-forms dual to the cycles labelled by $(A,I)$ in the previous section. The superpotential then reads 
\begin{equation}
    W = m_A \Pi^A + \tilde{m}_I\tilde{\Pi}^I\,. 
\end{equation}
In order to test the validity of the strong form of the tadpole conjecture introduced in section \ref{sec:review}, we take an approach to find solutions to $D_\alpha W=0$ first employed by \cite{Giryavets:2003vd} and subsequently used in e.g. \cite{Denef:2004dm,Cicoli:2013cha,Braun:2020jrx,Demirtas:2019sip}.
We therefore turn on flux only along $G$-invariant cycles, i.e.~$\tilde m_I = 0$, such that the superpotential reads 
\begin{equation}\label{Winv}
    W_{\rm inv} = m_A \Pi^A \,,
\end{equation}
where we used the basis of periods introduced in \eqref{def:periods} and chose and integer basis out of the $G$-invariant periods. Using the expansion of the $\Pi^A$ periods in \eqref{Aperiods} one finds 
\begin{equation}
\partial_{\phi^i} W|_{\{\phi^i=0\}} =0\,, \qquad \forall i\,,
\end{equation}
and using that the K\"ahler potential is at least quadratic in the $\phi^i$ one deduces
\begin{equation}\label{Kinull}
    \partial_{\phi^i} K|_{\{\phi^i=0\}}=0\,, \qquad \forall i\,. 
\end{equation}
Hence, along the symmetric locus, $\{\phi^i=0\}$, we automatically solve the F-term conditions for a superpotential of the form \eqref{Winv}.

Suppose we find a set of fluxes $m_A\neq 0$ for which we can also solve the remaining F-term constraints $D_{a}W =0$ at a generic point along the locus $\{\phi^i=0\}$. Let us further assume that $W_0\neq 0$. Given these assumptions, to determine the rank of $M$ in \eqref{defM} let us focus on the block corresponding to $(\alpha,\beta)=(i,j)$ which we split as
\begin{equation}\label{Msimplified}
    M = \underbrace{\left( \begin{matrix} D_i D_j W  & 0 \\ 0& \bar{D}_{\bar \imath} \bar{D}_{\bar \jmath} \bar W  \end{matrix} \right)}_{\equiv M_1}+ \underbrace{\left(\begin{matrix}
        0 & g_{\bar \imath j} W\\ g_{i \bar \jmath} \bar W& 0
    \end{matrix}\right)}_{\equiv M_2}\,.
\end{equation}
We saw in the previous section that the moduli space metric at a generic point along the locus $\{\phi^i=0\}$ is non-degenerate. From our discussion in section~\ref{sec:review} we recall that in the case $W_0\neq 0$ indeed half of the real directions in moduli space are obstructed automatically. From the perspective of the tadpole conjecture we are thus not interested in the full rank of $M$ but in $n_{\rm stab}$ as defined in \eqref{nstabAdS}. Using that the number of cancellations that can occur between $M_1$ and $M_2$ depends on the rank of $D_i D_j W$ we can give the following lower bound on $n_\mathrm{stab}$, 
\begin{equation}\label{deltalowerbound}
    n_\text{stab} \geq (h^{3,1}-h^{3,1}_{\rm inv}) - \mathrm{rank} (D_i D_j W )\,. 
\end{equation}
This bound does not yet use any assumption on the number of $G$-invariant moduli that are stabilized directly.
Assuming, however, that fluxes can be chosen in such a way that all $h^{3,1}_{\rm inv}$ invariant moduli obtain a mass, the bound \eqref{deltalowerbound} becomes  stronger,
\begin{equation} 
n_\text{stab}~\geq~h^{3,1}-\mathrm{rank}(D_iD_jW)\,.\end{equation}
In the regime $n_\text{stab} \gg 1$ the tadpole conjecture predicts that the $G_4$-flux induces a large tadpole bounded as in \eqref{tadpoleconj}.
To test this in the present setting, we thus require $h^{3,1}\gg 1$.
Luckily, the Calabi--Yau four-folds obtained as hyper-surfaces in weighted-projective space provide a large set of examples with large $h^{3,1}$. However, in order to achieve $n_\text{stab} \gg 1$ we still need to provide an upper bound on the rank of the Hessian $D_i D_j W$. 
Importantly, following \cite{Braun:2020jrx}, we can again exploit the presence of the symmetry $G$ to restrict $D_i D_j W$ considerably since the requirement of $G$-invariance enforces many elements of the matrix $D_i D_j W$ to vanish identically and hence often prohibits it from having full rank.

To see this a bit more concretely, recall that for our flux choice $\tilde m_I = 0$, and with the expansion \eqref{Aperiods} of the periods $\Pi^A$, the second covariant derivatives of the superpotential are given by
\begin{equation}
D_i D_j W = \pi^{A,2}_{ij}\left(m_A + \eta_{AB} \frac{\bar{\pi}^{B,0} \pi^{C,0}}{\left(K^{(0)}\right)^2} m_C \right)\,.
\end{equation}
Here, we used \eqref{Kinull} and from \eqref{OmegawedgebarOmega} deduced
\begin{equation}
    g_{ij}\big|_{\{\phi^i=0\}} = \frac{K_{ij}^{(2)}}{(K^{(0)})^2}\,. 
\end{equation}
Hence, a given element of the Hessian $D_i D_j W$ can only be non-vanishing if $\pi^{A,2}_{ij} \neq 0$ for at least one $A$.
On the other hand, since all $\Pi^A$ are $G$-invariant, a necessary condition for $\pi^{A,2}_{ij}$ to be non-zero is that 
\begin{equation}\label{invariance}
    \beta_i(g) \beta_j(g) = 1 \,,\qquad \forall g\in G\,,
\end{equation}
where the $\beta_i(g)$ characterize the transformation of the $\phi^i$ under $g\in G$ as in \eqref{transform}. In particular, we have 
\begin{equation}
    \beta_i(g_0) \beta_j(g_0) \neq 1\;\; \text{for some}\;g_0\in G \qquad \Rightarrow \qquad D_i D_j W \propto\pi^{A,2}_{ij}=0\,. 
\end{equation}
Moreover, we recall from the previous section that not all $\pi_{ij}^{A,2}$ are independent but related via the Jacobi-ideal $dP$ of $P$.
Therefore, moding out by $dP$ generates linear relationships between the elements of $D_i D_j W$ that potentially reduce its rank further.

\begin{figure}
    \centering
    \includegraphics[width=0.85\textwidth]{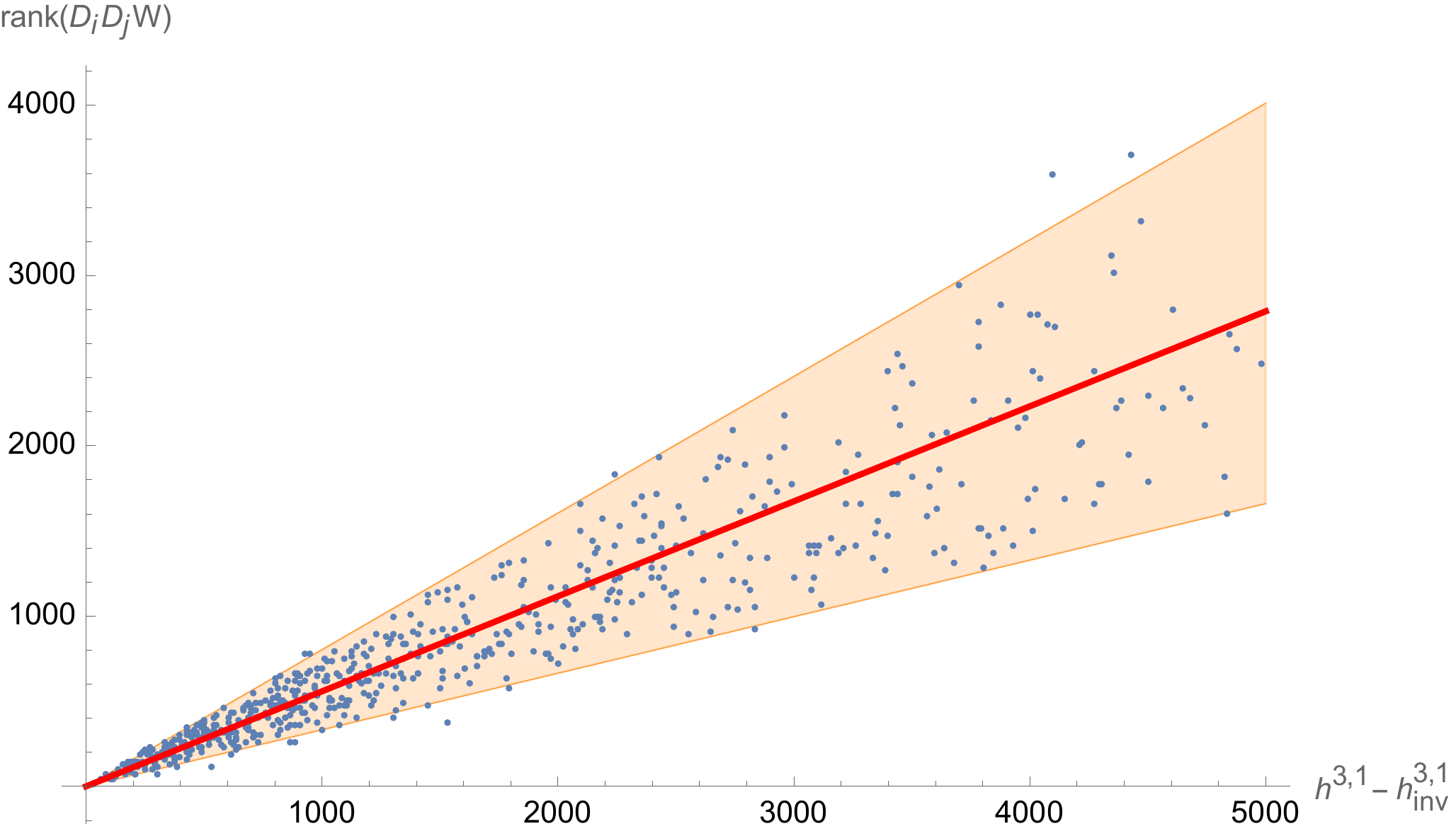}
    \caption{Shown is the maximal rank $r = \mathrm{rk} \, D_i D_j W$ of the Hessian of $W$ at a generic point along the locus $\{\phi^i=0\}$ in relation to $h^{3,1}-h^{3,1}_{\rm inv}$ for the cases with $h^{3,1}=h^{3,1}_{\rm poly}$. 95\% of the data points lie within the shaded region bounded by the two lines with slope $r=0.80\,(h^{3,1}-h^{3,1}_{\rm inv})$ and $r = 0.33\,(h^{3,1}-h^{3,1}_{\rm inv})$. The red, solid line indicates the mean behavior of $r \sim 0.56\, (h^{3,1}-h^{3,1}_{\rm inv})$.}
    \label{fig:rankM1}
\end{figure}

For a given Fermat-type Calabi-Yau hypersurface with known symmetry group $G$ and group actions $\beta_i(g)$, the invariant combinations $(i,j)$ for which \eqref{invariance} is satisfied can be determined explicitly.
This yields all elements of $D_i D_j W$ that are not identically zero by symmetry reasons.
Assuming that these elements are all non-vanishing and independent, and taking into account the linear relations imposed by the Jacobi-ideal, one can use this information to compute the maximal possible rank of $D_i D_j W$.
We do this systematically using the computer algebra system Mathematica for a large number of Fermat-type Calabi-Yau four-folds in our dataset.%
\footnote{Our Mathematica code and data is publicly available at \cite{FermatCY}.}
Hereby, we restrict to  hypersurfaces with $h^{3,1}=h^{3,1}_{\rm poly}$ and  $h^{3,1} - h^{3,1}_\mathrm{inv} < 5000$, resulting in a dataset of 609 different Calabi-Yau four-folds%
\footnote{We furthermore considered only the first 3000 Fermat-type four-folds and restricted each computation to maximally 8GB of memory.
Therefore, there are a few Calabi-Yau four-folds with $h^{3,1} - h^{3,1}_\mathrm{inv} < 5000$ that are not contained in our list.}
The results are illustrated in figure \ref{fig:rankM1}.  

Let us now denote the rank of the matrix $D_i D_j W$ by $r$.
We are particularly interested in the scaling behavior of  $r$ with respect to the number of non-invariant moduli $h^{3,1}-h^{3,1}_{\rm inv}$.
From figure~\ref{fig:rankM1} one sees that $r$ grows approximately linear in $h^{3,1}-h^{3,1}_{\rm inv}$. More precisely, we find 
\begin{equation}
 r \sim \gamma \left(h^{3,1}-h^{3,1}_{\rm inv}\right) \,, \qquad \text{with} \qquad  \gamma\in [0.33,0.80] \,.
\end{equation}
The latter bound on $\gamma$ is satisfied for 95\% of all Fermat-type four-folds we considered and corresponds to the shaded area in figure~\ref{fig:rankM1}.
We can use this bound and \eqref{deltalowerbound} to give a lower-bound on the number of stabilized moduli,
\begin{equation}\label{bound}
    n_{\rm stab} \geq  (1-\gamma) \,  (h^{3,1}-h^{3,1}_{\rm inv})\gtrsim 0.2\,  (h^{3,1}-h^{3,1}_{\rm inv})\,. 
\end{equation}
Accordingly, just by turning on a flux in the invariant directions, we can achieve a large number of stabilized moduli as long as $W_0\neq 0$. 
Notice that this bound is saturated if all cancellations between the two matrices in \eqref{Msimplified}, that are allowed by $G$-invariance, are actually realized.
It is conceivable that in the generic case the number of such cancellations is considerably smaller or even zero, resulting in a much larger number of stabilized moduli than required by this bound.

Even though here we do not attempt to calculate the exact physical masses of the moduli, let us remark that the mass of the scalar deformations that lie the kernel of $D_i D_j W$ is set by the value of $W_0$ whereas for the rest the actual mass depends also on $D_i D_j W$ and thus on the form of $\pi^{A,2}_{ij}$.\footnote{This implies that the mass of the scalar deformation modes is not universal, but depends on the exact form of the periods in the vicinity of the symmetric locus. This should be contrasted with the results of \cite{Blanco-Pillado:2020hbw,Blanco-Pillado:2020wjn} where it was found that the non-invariant deformations have a universal mass. To our understanding the difference is that \cite{Blanco-Pillado:2020hbw,Blanco-Pillado:2020wjn} use the LCS expressions also for the non-invaraint periods which is not justified as the symmetric locus is far away from the LCS phase.} On the other hand, the fermion mass matrix is given simply by $D_iD_jW$ such that we expect at least $0.2\, (h^{3,1}-h^{3,1}_{\rm inv})$ massless fermions. 

Let us finally contrast the $W_0\neq 0$ situation with the case in which we find a supersymmetric vacuum $W_0=0$ along the symmetric locus $\{\phi^i=0\}$. As discussed in section~\ref{sec:review}, in this case we should indeed count the number of stabilized complex scalar fields, i.e. use the definition of $n_{\rm stab}$ in \eqref{nstabMink}. For $W_0=0$, the matrix $M_2$ does not provide an obstruction for any complex structure deformations. Therefore any first-order obstruction needs to arise from $D_i D_j W$. In this case we simply have 
\begin{equation}\label{boundW=0}
    n_{\rm stab}\Big|_{W_0=0} = \;\text{rk}\,(D_i D_j W) \lesssim 0.8 \,(h^{3,1}-h^{3,1}_{\rm inv})\,. 
\end{equation}
Notice that this bound assumes that all entries of $D_iD_jW$, that can be non-zero by symmetry considerations, are indeed non-vanishing. Thus, whereas for cases with $W_0\neq 0$ we can give a lower bound on $n_{\rm stab}$, for $W_0=0$ we only have an upper bound on $n_{\rm stab}$ for invariant fluxes. This matches with the results of \cite{Braun:2020jrx,Becker:2022hse} where it was found that in general supersymmetric vacua at the Fermat point obtained from symmetric fluxes have a large number of flat directions.

\subsection{Consequences for the tadpole conjecture}
Given the bound on the co-dimension $n_{\rm stab}$ of $\mathcal{M}_{\rm vac}$ we now want to discuss the consequences for the tadpole conjecture of \cite{Bena:2020xrh} reviewed and refined in section~\ref{sec:review}. As we discussed in the previous section, the fourfold $Y_4$ is in general smooth along the symmetric locus unless we tune the $\psi^a$ to special values. Therefore at a generic point along the symmetric locus we meet the requirements of both, the strong and the weak version of the tadpole conjecture (as introduced above).
In its stronger form this conjecture predicts a bound as in \eqref{tadpoleconj} relating $n_{\rm stab}$ and the M2-brane tadpole induced by the choice of flux. To check the validity of this statement in our setup we thus need to give an estimate for the tadpole generated by turning on invariant fluxes $m_A$. As alluded to before, in favorable cases this is equivalent to turning on generic fluxes along the mirror $X_4$ and our problem thus reduces to finding self-dual fluxes for $X_4$ with $W_0\neq 0$ and relatively small $N_{\rm flux}$. 

Since typically $h^{3,1}_{\rm inv} \ll h^{3,1}$ this is a much simpler task. In particular if $h^{3,1}\lesssim  \mathcal{O}(10)$ this can be done explicitly for instance in the limit $\psi^a\rightarrow \infty$, i.e., the large complex structure regime, using, e.g., the expressions derived in \cite{Marchesano:2021gyv}. To find solutions to $D_a W=0$ for such a small number of moduli in this regime, we expect that it is sufficient to turn on a flux with $N_{\rm flux}\sim \mathcal{O}(10)$. In particular, the strong version of the tadpole conjecture itself is consistent with such an expectation. Let us therefore \emph{assume} that there exists a bound of the form \eqref{tadpoleconj} for the $h^{3,1}_{\rm inv}$ parameters that can marginally be satisfied. The tadpole conjecture would then require
\begin{equation}\label{relation}
 \alpha\, h^{3,1}_{\rm inv} \stackrel{?}{\geq} (1-\gamma)(h^{3,1}-h^{3,1}_{\rm inv})\,.
\end{equation}
For $\alpha\sim  \mathcal{O}(1)$ we notice that this cannot be satisfied for $h^{3,1}\gg h^{3,1}_{\rm inv}$. Among the Fermat-type hypersurfaces there are indeed examples with $h^{3,1}/h^{3,1}_{\rm inv}\sim \mathcal{O}(1000)$ implying that \eqref{relation} does not hold in general.
We therefore conclude that on the symmetric locus of $Y_4$ the strong version of the tadpole conjecture might not hold in general.
To see how this works explicitly, we present a simple example in the next section. 

As reviewed in section~\ref{sec:review}, a weaker form of the tadpole conjecture only states that there is an obstruction to stabilize \emph{all} complex structure moduli within the tadpole bound, leading to the bound \eqref{tadpoleconj2}. In order to apply our results to this case, we need to ensure that if the rank of the matrix $M_1$ along the locus $\{\phi^i=0\}$ is non-zero, there are no cancellations between $M_1$ and $M_2$ leading to a reduction of the rank of $M$ altogether. To check for this possibility, let us compare the two matrices at the locus $\{\phi^i=0\}$: 
\begin{equation}
\begin{aligned}
    (M_1)_{i,j} &= \pi^{A,2}_{ij}(\psi^a)\left(m_A + \eta_{AB} \frac{\bar{\pi}^{B,0}(\bar{\psi}^a) \pi^{C,0}(\psi^a)}{\left(K^{(0)}(\psi^a,\bar{\psi}^b)\right)^2}m_C\right)\,,\\ (M_2)_{i\bar \jmath} &= \frac{K^{(2)}_{i\bar{\jmath}}(\psi^a,\bar{\psi}^b)}{K^{(0)}(\psi^a,\bar{\psi}^b)^2} \left[m_A \pi^{A,0}(\psi^a)\right]\,. 
    \end{aligned}
\end{equation}
For generic $\psi^a$ and fluxes $m_A$ we do not expect a cancellation between these two matrices. Still, this might be the case for the special values of $\psi^a$ for which $D_a W=0$. We notice, however, that solving $D_aW=0$ along the locus $\{\phi^i=0\}$ is completely independent of the form of the functions $\pi^{A,2}_{ij}$ and $K^{(2)}_{i\bar\jmath}$. We therefore do not expect that along the locus $D_aW=0$ these two functions in general conspire to lead to a decrease of the rank of $M$.\footnote{In contrast to that, in the next section we discuss how such a cancellation can occur in asymptotic limits.} We therefore expect that indeed the rank of $M$ is $h^{3,1}$ if $W_0\neq 0$. To scrutinize this expectation we would need to have the exact $\psi^a$-dependence of the relevant functions for general $\psi^a$ at our disposal. To date the exact form of these functions is, however, not known for general CYs. Therefore, at this stage we cannot conclusively show that full moduli stabilization of a large number of moduli at the symmetric point is possible even though our argument indicates that there should be no fundamental obstruction to achieve that. Whereas the strong form of the tadpole conjecture turns out not be realized at the symmetric locus, it is hence still possible that the weaker form holds. 

\subsection{Example: \texorpdfstring{$\mathbb{P}^5_{1,1,1,1,8,12}[24]$}{P5(1,1,1,1,8,12)[24]}}\label{sec:ex}
Let us consider a Calabi--Yau fourfold $Y_4$ obtained as the degree-24 hypersurface in $\mathbb{P}^5_{1,1,1,1,8,12}$ determined by the zero locus of the polynomial in Fermat form
\begin{equation}\label{eq:CY1111812}
    P_0=x_1^{24}+x_2^{24}+x_3^{24} +x_4^{24}+x_5^3+x_6^2=0\,.
\end{equation}
Na\"ively, the resulting Calabi-Yau hypersurface is invariant under a $\tilde G=\mathbb{Z}_{24}^3\times \mathbb{Z}_3 \times \mathbb{Z}_2$ symmetry.
The generators of $\tilde{G}$ can be found by solving \eqref{generatorcond}, with a possible choice given by
\begin{equation}\begin{gathered}\label{eq:generators}
    g^{(1)} =(-1,1,0,0,0,0)\,,\quad g^{(2)}=(-1,0,1,0,0,0)\,,\quad g^{(3)}=(-1,0,0,1,0,0) \,, \\
    g^{(4)} =(-8,0,0,0,1,0) \,,\quad g^{(5)} =(-12,0,0,0,0,1 ) \,,
\end{gathered}\end{equation}
acting on the coordinates of $\mathbb{P}^5_{1,1,1,1,8,12}$ as in \eqref{gaction}.
However, when taking the rescaling invariance \eqref{eq:lambdascaling} of the projective coordinates into account, these five generators are not all independent.
For example, the following combinations can be seen to act trivially
\begin{equation}\label{eq:trivialgenerators}
\left[ g^{(1)} g^{(2)} g^{(3)} \right]^4  g^{(4)} =
\left[ g^{(1)} g^{(2)} g^{(3)} \right]^3  g^{(5)} = 1 \,,
\end{equation}
implying that $g^{(4)}$ and $g^{(5)}$ are not independent but can be expressed in terms of the first three generators.
Moreover, squaring the second equation in \eqref{eq:trivialgenerators} gives $\left[ g^{(1)} g^{(2)} g^{(3)} \right]^6 = 1$, reducing one of the $\mathbb{Z}_{24}$ factors to $\mathbb{Z}_6$. 
Therefore, the actual symmetry of the Calabi-Yau defined in \eqref{eq:CY1111812} is
\begin{equation}
G=\mathbb{Z}_{24}^2\times\mathbb{Z}_6 \,,
\end{equation} 
and is spanned by the first three generators $g^{(1)}$, $g^{(2)}$, and $g^{(3)}$ in \eqref{eq:generators}.

In total the hypersurface $\{P_0=0\}$ allows for 3878 polynomial complex structure deformations. 
This can be seen as follows:
The space of independent deformations is spanned by all degree-24 polynomials in the variables $x_1, \dots, x_6$, divided by the Jacobian ideal $\left<\partial P_0\right> = \left<x_1^{23}, x_2^{23},x_3^{23},x_4^{23},x_5^2, x_6\right>$.
A suitable basis is hence given by the monomials
\begin{equation}\label{eq:indepdefs}
x_1^{\alpha_1} x_2^{\alpha_2} x_3^{\alpha_3} x_4^{\alpha_4} \,,\qquad x_1^{\beta_1} x_2^{\beta_2} x_3^{\beta_3} x_4^{\beta_4} x_5 \,,
\end{equation}
where $\sum_i \alpha_i = 24$, $\alpha_i < 23$ and $\sum_i \beta_i = 16$.
Since the number of degree-$p$ monomials in $n+1$ variables is  $\begin{pmatrix} p+n \\ n \end{pmatrix}$ we find that the total number of deformations is given by
\begin{equation}
\begin{pmatrix}27 \\ 3  \end{pmatrix} - 4 - 4 \times 3 + \begin{pmatrix}19 \\ 3  \end{pmatrix} = 3878 \,,
\end{equation}
where we accounted for the over-counting of monomials that violate the condition $\alpha_i < 23$.

Resolving the orbifold singularities induced by the $\mathbb{C}^*$-action of the projective space introduces an exceptional divisor $E$ but no further complex structure deformations. We thus have $(h^{1,1},h^{3,1})=(2,3878)$. Out of the complex structure deformations there are two deformations which respect the discrete symmetry.
Using the generators $g^{(1)}$, $g^{(2)}$, and $g^{(3)}$ of $G = \mathbb{Z}_{24}^2\times \mathbb{Z}_6$ in \eqref{eq:generators}, one sees that they are of the form \eqref{eq:indepdefs} with all powers of $x_1, \dots, x_4$ equal.
Therefore, a degree-24 hypersurface respecting the discrete symmetry takes the form 
\begin{equation}
    P_{\rm inv.}(\psi^0, \psi^1)= x_1^{24}+x_2^{24}+x_3^{24} +x_4^{24}+x_5^3+x_6^2-\psi^0 x_1^6x_2^6x_3^6x_4^6 -\psi^1 x_1^4x_2^4x_3^4x_4^4x_5=0\,,
\end{equation}
where we only kept terms modulo the ideal $\langle \partial P_0\rangle$.
Restricting the 3878-dimensional moduli space to the symmetric locus, where we set all polynomial deformations of $P$ not invariant under $G$ to zero, we are effectively left with a two-dimensional moduli space spanned by $(\psi^0, \psi^1)$. Notice that the hypersurface $\{P=0\}\subset \mathbb{P}^5_{1,1,1,1,8,12}$ can be thought of as a smooth elliptic fibration over $\mathbb{P}^3$.\footnote{Since it is an elliptic fibration, it allows for an uplift to F-theory such that the results we obtain in this appendix can also be lifted to the 4d case and are not a special effect of the underlying 3d M-theory.} Since $Y_4$ has $h^{1,1}=2$, its mirror $X_4$ has $h^{3,1}=2$ such that we can treat the symmetric locus spanned by $(\psi^0,\psi^1)$ as the full complex structure moduli space of $X_4$ or, equivalently, as the K\"ahler moduli space of $Y_4$.

We take this latter perspective and analyze moduli stabilization in the large volume/large complex structure regime where the mirror map identifies the complexified K\"ahler moduli of $Y_4$ can be identified with the invariant deformations $(\psi^0,\psi^1)$ as 
\begin{equation}
    b^a+i t^a = -\frac{1}{2\pi i }\log(\psi^a) +\dots \,,\quad \text{for}\;\; \psi^a\rightarrow \infty\,,\quad a=0,1\,. 
\end{equation}
The dots indicate terms polynomial in $1/\psi^a$ that are suppressed in the large complex structure limit. This two-dimensional moduli space has already been investigated in some detail in \cite{Cota:2017aal,Marchesano:2021gyv} in the context of flux vacua. Let us review some details. The K\"ahler cone of $Y_4$ is generated by two divisor classes $D_0,D_1$ which can be identified as 
\begin{equation}
    D_0= E + 4 \pi^*H \,,\qquad D_1 = \pi^* H\,,
\end{equation}
where $E$ is the exceptional divisor, i.e. the zero section of the elliptic fibration $\pi: T^2\rightarrow \mathbb{P}^3$ and $H$ the hyperplane class of $\mathbb{P}^3$. Their intersection numbers can be summarized in the intersection polynomial
\begin{equation}
    I(Y_4)= 64D_0^4 +16D_0^3D_1 +4D_0^2D_1^2 + D_0D_1^3\,. 
\end{equation}
The dual Mori cone generators $\mathcal{C}^{0,1}$ can accordingly be identified with the generic elliptic fiber and the single Mori cone generator of $\mathbb{P}^3$, respectively. In addition, we have two classes of four-cycles in $Y_4$ 
\begin{equation}
    [H_1] = [D_0.D_1]\,,\qquad [H^1]=\pi^*[\mathcal{C}^1]\,. 
\end{equation}
Together with the point class and the class of the full Calabi--Yau we thus get the K-theory basis 
\begin{equation}
    (\mathcal{O}_{\rm pt}, \iota_!\mathcal{O}_{\mathcal{C}^0}(K_{\mathcal{C}^0}^{1/2}),  \iota_!\mathcal{O}_{\mathcal{C}^1}(K_{\mathcal{C}^1}^{1/2}), \mathcal{O}_{H^1}, \mathcal{O}_{H_1},, \mathcal{O}_{D_1},\mathcal{O}_{D_0}, \mathcal{O}_{X_4})\,,
\end{equation}
where $\mathcal{O}_{Y_4}$ is the structure sheaf on $Y_4$. Via mirror symmetry we can identify this basis with an integer basis of fluxes on $X_4$ which, following \cite{Marchesano:2021gyv}, we denote by 
\begin{equation}
    (e, e_0, e_1, m_1, \hat{m}^1, m^1,m^0, m)\,. 
\end{equation}
In order to apply the analysis of \cite{Marchesano:2021gyv} in the large structure regime, we also need to specify some additional data of $Y_4$ such as the Chern classes 
\begin{equation}
    K^{(2)}_{ij} = \frac{1}{24} \int_{Y_4} c_2(Y_4)\wedge D_i\wedge D_j\,,\qquad K_i^{(3)} = \frac{\zeta(3)}{8\pi^3}\int_{Y_4} c_3(Y_4)\wedge D_i\,,
\end{equation}
which in the present case are given by
\begin{equation}\begin{aligned}
    K_{00}^{(2)} &= \frac{91}{3} \,,\quad K_{01}^{(2)} = K_{10}^{(2)}= \frac{91}{12}\,,\quad K_{11}^{(2)}=2\,,\\ K_0^{(3)}&=-3860 \frac{\zeta(3)}{(2\pi)^3} \,,\quad K_1^{(3)}=-960 \frac{\zeta(3)}{(2\pi)^3}\,. 
\end{aligned}\end{equation}
Notice that $c_2(Y_4)$ is even such that we do not need to introduce a vertical flux in order to satisfy the flux-quantization condition \eqref{eq:quantization}. We want to find solutions to $D_0W=D_1W=0$ in the large complex structure regime. As discussed in \cite{Marchesano:2021gyv} to find such solutions while keeping the tadpole relatively small, we should set $m^0=m^1=m=0$. In this case the tadpole is simply given by 
\begin{equation}
    N_{\rm flux} = \hat{m}^1(m_1+2\hat{m}^1)\,. 
\end{equation}
The F-term conditions now amount to \cite[eqs.~(6.44), (6.46)]{Marchesano:2021gyv}  which can be solved for $t^0$ and $t^1$. For consistency of the LCS approximation we require that the solutions for $t^i$ lie sufficiently deep inside the LCS phase. For instance taking the flux choice
\begin{equation}
    (e,e_0,e_1,m_1,\hat{m}^1, m^1, m^0, m) = (-61,4,15,1,1,0,0,0)\,,
\end{equation}
we obtain a solution to the F-terms at 
\begin{equation}
   (t^0,t^1) \sim (5.18, 16.77)\,,
\end{equation}
which lies within the region where we can approximate the actual moduli space geometry by its LCS expressions.\footnote{To see this, we note that these values for the $t^i$ correspond to values for $|\psi^i|\gg 10^{10}$. Since there are no singular loci for $\psi^0>864$ and $\psi^1 > 256$ (cf. for instance \cite{Cota:2017aal}) at which the LCS approximation breaks down, we are well inside the regime of applicability of the expressions of \cite{Marchesano:2021gyv}.} This flux choice contributes to the tadpole as
\begin{equation}
    N_{\rm flux}=3\,,
\end{equation}
which is well below the available tadpole
\begin{equation}
    \frac{\chi(Y_4)}{24}= 972\,.
\end{equation}
On the other hand, the on-shell value of $W$ is given by
\begin{equation}
    |W_0|= \frac12 \left[\hat{m}^1 (t^1)^2 + m_1\left(2t^0t^1 +4(t^0)^2\right)\right]\sim 237.79\neq 0\,. 
\end{equation}
Following our general discussion in this section, we can thus expect that through this flux $n_{\rm stab}$ is bounded by
\begin{equation}\label{eq:appnstabbound}
    n_{\rm stab} \geq  h^{3,1} - \text{rk} \,(D_i D_j W)\,,
\end{equation}
where we used that the chosen flux stabilizes all invariant moduli $\psi^{0,1}$.

To find the elements $D_i D_j W$ that are not identically zero, we have to determine all products of two monomials of the form \eqref{eq:indepdefs} that are invariant with respect to $G$. 
After dividing by the Jacobian ideal $\left<\partial P\right>$, there are only two such independent, rank-48 monomials, 
\begin{equation}
x_1^{12} x_2^{12} x_3^{12} x_4^{12} \,,\quad\text{and}\quad x_1^{10} x_2^{10} x_3^{10} x_4^{10} x_5 \,,
\end{equation}
reflecting the fact that $h_\mathrm{inv}^{2,2} = 2$.
Importantly, while the number of independent polynomials is independent of the point at which $\left<\partial P\right>$ is evaluated, it does play a role for the computation of $D_i D_j W$.
For example, at the Fermat point, i.e. $\psi^a  = 0$, all monomials containing $x_5^2$ are elements of $\left<\partial P\right>$ and the corresponding elements of $D_i D_j W$ vanish identically.
On the other hand, at a generic point in moduli space, $\psi^a \neq 0$, dividing by $\left<\partial P\right>$ creates a linear relationship between such monomials and other terms but does not set them to zero.
Therefore, away from the Fermat point $D_i D_j W$ can have generically additional non-vanishing entries.

Using our Mathematica code we can go through all the possible combinations of monomials of the form \eqref{eq:indepdefs}.
We find that there are 2956 $G$-invariant combinations at the Fermat point, while at a generic point on the symmetric locus this number is increased to 5986.
We furthermore compute the rank of the resulting matrix $(D_i D_j W)$, assuming that all its elements are non-vanishing, and find that it is bounded by 2824.
Therefore, according to our formula \eqref{eq:appnstabbound}
this flux choice leads to 
\begin{equation}
    n_{\rm stab}\geq 1052\,,
\end{equation}
while inducing a tadpole of just $N_{\rm flux}=3$. This example thus shows explicitly that at the symmetric locus we can achieve
\begin{equation}
    \frac{n_{\rm stab}}{N_{\rm flux}} \geq \frac{1052}{3} \gg 1 \,,
\end{equation}
which violates \eqref{tadpoleconj} indicating that the strong form of the tadpole conjecture is not valid at the symmetric locus.

\section{Comparison to other regimes}\label{sec:comparison}
The validity of the tadpole conjecture has previously been under investigation in other regimes of the complex structure deformation space such as the large complex structure regime \cite{Marchesano:2021gyv,Plauschinn:2021hkp,Lust:2021xds,Grimm:2021ckh,Tsagkaris:2022apo} or even more general asymptotic limits \cite{Grana:2022dfw}. The hall-mark of these limits is that there is a unipotent monodromy around an asymptotic locus in moduli space that can be exploited to analyze the moduli dependence of the Hodge star operator involved in the self-duality constraint $G_4 = *G_4$. This monodromy can be associated to a continuous shift-symmetry that is approximately realized in the vicinity of the asymptotic locus. The shift symmetry can be made explicit by splitting the complex structure fields into a saxionic and an axionic component which is reflected in the asymptotic form of the periods. By contrast around the symmetric locus $\{\phi^i=0\}$ of the families of fourfolds considered in this work, we do not have such a shift-symmetry. Instead, the form of the periods in the vicinity of the locus $\{\phi^i=0\}$ is now constrained by the discrete symmetry $G$. Hence, the loci considered so far in this work do not fall in the class of asymptotic loci considered in \cite{Grana:2022dfw}.

Given the decomposition of the primitive middle cohomology of $Y_4$ as 
\begin{equation}
 H^4 = H^{4,0} \oplus H^{3,1} \oplus H^{2,2}_{\rm prim} \oplus H^{1,3} \oplus H^{0,4}\,,
\end{equation}
self-duality of the flux implies that at the critical point $D_a W=0$ the $G_4$-flux cannot have any $(3,1)$ and $(1,3)$ components. Since $W= \int \Omega \wedge G_4$, the condition $W_0\neq 0$ further implies that $G_4$ has a component in $H^{0,4}$. In \cite{Grana:2022dfw} it was shown that if one wants to solve the F-term condition in strict asymptotic limits using the leading expression for the periods, it is almost irrelevant whether $G_4$ has a $(0,4)$ component or not in order to achieve $n_{\rm stab}\gg 1$. In these limits most of the directions in complex structure moduli are obstructed by fluxes that are exclusively in $H^{2,2}$. More precisely, the $(0,4)$ component of $G_4$ can only obstruct up to four directions in the complex structure moduli space. This is in stark contrast to our findings for the case of the symmetric locus: We find that a large number of non-invariant complex structure deformations are obstructed merely by the presence of a $(0,4)$ component of $G_4$ implying a non-zero $W_0$. The comparison to the asymptotic regimes investigated in \cite{Grana:2022dfw} thus nicely illustrates that the physics in asymptotic regimes can differ significantly from the physics in the deep interior of moduli space. 

In order to understand this difference better, let us consider the matrix $M$ in \eqref{defM} in an asymptotic regime with approximate continuous shift symmetry. In order to avoid confusion, let us denote the complex scalar fields in this regime by $T^\alpha$, $\alpha=1, \dots, h^{3,1}$, on which the shift symmetry acts as $T^\alpha \rightarrow T^\alpha + c$ for $c\in \mathbb{R}$. In case we have such a shift symmetry, the K\"ahler potential is a function of the imaginary part of $T^\alpha$, i.e.
\begin{equation}\label{Kahlershiftsymm}
    K(T^\alpha, \bar T^\alpha) = K(T^\alpha-\bar T^\alpha)\,. 
\end{equation}
Again, $M_2$ as in \eqref{Msimplified} naively leads to an obstruction for all complex structure deformations. However, $D_\alpha D_\beta W$ contains a term 
\begin{equation}
    D_\alpha D_\beta W \ni g_{\alpha \beta} W\,,
\end{equation}
with $ g_{\alpha \beta}=\partial_{T^\alpha}\partial_{T^\beta}K$. Due to the shift symmetry we have $g_{\alpha \beta}= -g_{\alpha \bar \beta}$ such that 
\begin{equation}\label{matrixgij}
    \text{ker}\,\begin{pmatrix}  g_{\alpha \beta} W & g_{\bar \alpha \beta }W \\ g_{\alpha \bar \beta}\bar{W}& g_{\bar\alpha \bar \beta} \bar{W}
    \end{pmatrix}=\text{span}_{\alpha=1,\dots, h^{3,1}}\langle v^\alpha\rangle \,,
\end{equation}
with 
\begin{equation}
v^\alpha=(0,\dots, 0,\text{Re}\,T^\alpha,0,\dots 0; 0, \dots,0, \text{Re}\,T^\alpha, 0, \dots, 0)\,.
\end{equation} 
Hence, this contribution to $D_\alpha D_\beta W$ cancels half of the obstructions induced by $M_2$. In particular, at this point there is no obsturction to changing the value of the axionic component of $T^\alpha$. On the other hand, the saxionic directions of $T^\alpha$ naively are obstructed unless the metric $g_{\alpha \bar \beta}$ is degenerate. This reflects the fact that for $W_0\neq 0$ half of the real moduli are automatically stabilized and the residual moduli space is real and of maximal dimension $h^{3,1}$ \cite{deAlwis:2013jaa}. However, the results of \cite{Grana:2022dfw} imply that at a point away from the singular locus there are further cancellations between $M_1$ and $M_2$ reducing the rank of $M$ further unless the flux is chosen suitably to have enough components in the different subspaces of $H^{2,2}$. In particular this implies that $D_\alpha D_\beta W$ also has almost full rank allowing for such a cancellation. This is precisely the difference to the symmetric locus analyzed in the previous sections where we showed that $M_1$ in general has rank considerably smaller than $h^{3,1}$ therefore allowing us to infer that many additional directions in the field space are automatically obstructed just by the non-vanishing of $W_0$. 

While we do not attempt to show how this cancellation works in the general asymptotic considered in \cite{Grana:2022dfw} let us briefly discuss this cancellation in one of the large complex structure scenarios discussed in \cite{Marchesano:2021gyv}: In the large complex structure regime, the K\"ahler potential \eqref{Kahlershiftsymm} can be conveniently expressed in terms of the intersection numbers $\mathcal{K}_{\alpha \beta \gamma \delta}$ of the mirror as 
\begin{equation}
    K=-\log(\frac{2}{3}\mathcal{K}_{\alpha \beta \gamma \delta}t^\alpha t^\beta t^\gamma t^\delta)\,.
    \label{Kcs}
\end{equation}
Here we split $T^\alpha=b^\alpha+it^\alpha$. Hence, $g_{\alpha \beta}=-g_{\alpha \bar \beta}$ and the rank of the matrix in \eqref{matrixgij} is indeed reduced by a factor of two. In particular, this contribution to $M$ now only gives a mass to the imaginary part of $T^\alpha$, i.e.\ the saxionic part. Let us focus on the generic scenario of \cite{Marchesano:2021gyv}. In this case the superpotential is given by 
\begin{equation}
 W = \bar{e} +\bar{e}_\alpha T^\alpha +\frac12 \zeta_{\mu, \alpha \beta} m^\mu T^\alpha T^\beta\,, 
\end{equation}
where $\zeta_{\mu, \alpha \beta}$ is a tensor that depends on the topological data of the CY fourfold and $\bar{e}$, $\bar{e}_\alpha$ and $m^\mu$ are related to quantized fluxes in a certain basis. As shown in \cite{Marchesano:2021gyv} the condition $D_\alpha W=0$ enforces (without taking into account any corrections to the leading order periods)
\begin{equation}\label{eom}
    \bar{e} = -\frac12 e_\alpha b^\alpha \,,\qquad e_\alpha = -\zeta_{\mu, \alpha\beta} m^\mu b^\beta\,,\qquad \left(\mathcal{K} \zeta_{\mu, \alpha} - \mathcal{K}_\alpha \zeta_\mu \right)m^\mu=0\,. 
\end{equation}
Here we introduced $\mathcal{K}_\alpha=\mathcal{K}_{\alpha \beta \gamma \delta}t^\beta t^\gamma t^\delta$ and $\mathcal{K}=t^\alpha\mathcal{K}_\alpha$, as well as $\zeta_{\mu, \alpha}=\zeta_{\mu,\alpha\beta}t^\beta$ and $\zeta_{\mu}=\zeta_{\mu,\alpha\beta}t^\alpha t^\beta$ . Accordingly, on-shell we find 
\begin{equation}
    W_0 = -\frac12  \zeta_{\mu, \alpha \beta} m^\mu t^\alpha t^\beta\,.
\end{equation}
The matrix $M$ is now given by 
\begin{equation*}
 M =\left(\zeta_{\mu,\alpha \beta}m^\mu - \frac{2\mathcal{K}_\alpha\mathcal{K}_\beta}{\mathcal{K}^2}\zeta_{\mu, \gamma \delta} m^\mu t^\gamma t^\delta \right)\otimes \left( \begin{matrix}  1 & 0 \\ 0& 1  \end{matrix} \right) -\frac{ \zeta_{\mu,\gamma \delta} m^\mu t^\gamma  t^\delta }{2}  \left(\begin{matrix}
        g_{\alpha \beta }  & g_{\bar \alpha \beta} \\ g_{\alpha \bar \beta} & g_{\bar \alpha \bar \beta}
    \end{matrix}\right)\,,
\end{equation*}
with
\begin{equation}
    g_{\alpha\beta } = -g_{\alpha \bar \beta} = -4 \frac{\mathcal{K}_\alpha \mathcal{K}_\beta}{\mathcal{K}^2} + 3 \frac{\mathcal{K}_{\alpha \beta}}{\mathcal{K}}\,. 
\end{equation}
Since the second term in $M$ does not stabilize the real part of $T^\alpha$, the obstruction for these axionic directions comes entirely from the first part. And indeed the rank of the first term in $M$ is essentially determined by the rank of $\zeta_{\mu, \alpha \beta}m^\mu$ in accordance with the analysis of \cite{Marchesano:2021gyv}. On the other hand, generically we expect all saxions $t^\alpha$ to get a mass from the second term in $M$ unless there is a cancellation. And indeed we find that there is a flat direction since 
\begin{equation}
 M \left(\begin{matrix} t^\alpha \\ -t^\alpha \end{matrix}\right) =0\,. 
\end{equation}
To see this, we calculate explicitly
\begin{equation}
 \zeta_{\mu, \beta} m^\mu - 2 \frac{\mathcal{K}_\beta}{\mathcal{K}} \zeta_\mu m^\mu +  \frac{\mathcal{K}_\beta}{\mathcal{K}} \zeta_\mu m^\mu = \frac{1}{\mathcal{K}}\left(\mathcal{K} \zeta_{\mu,\beta} -\mathcal{K}_\beta\zeta_\mu \right)m^\mu =0\,,
\end{equation}
where in the last step we used \eqref{eom}. Hence, in this approximation there is indeed a flat direction which gets lifted by including polynomial corrections to the periods. Thus, including corrections we do not expect any cancellations between the two contributions, $M_1$ and $M_2$, to $M$ anymore. As discussed in the previous section also for the case of stabilization on the symmetric locus we do not expect such a cancellation between $M_1$ and $M_2$ in general. 
\section{Discussion}\label{sec:conclusions}
In this note we investigated moduli stabilization and the tadpole conjecture \cite{Bena:2020xrh} in the interior of the complex structure moduli space of M-theory flux-compactifications on Calabi--Yau fourfolds.  We suggested to differentiate between a strong and a weak version of the tadpole conjecture.
The strong version states that stabilizing any number $n_{\rm stab} \leq h^{3,1}$ of directions in moduli space requires $G_4$-flux with an M2-brane tadpole that is directly proportional to $n_{\rm stab}$.
On the other hand, the weak version states that this only applies to the case $n_{\rm stab}=h^{3,1}$ implying that there is only a problem to achieve  \emph{full} moduli stabilization for $h^{3,1}\gg 1$.  Whereas there is evidence \cite{Grana:2022dfw} that the strong tadpole conjecture is satisfied in strict asymptotic limits in the vicinity of infinite distance points, it is a priori not clear how these result extend to the interior of the moduli space.

In this work, we mostly focused on the case of $W_0\neq 0$ vacua. In this case, half of the real deformations are automatically stabilized.
Therefore, we proposed that the number $n_{\rm stab}$ in the tadpole conjecture should only count \emph{additionally} stabilized real directions in moduli space.
Following this convention, we proceeded to test the strong version of the tadpole conjecture in the interior of the moduli space. To that end, we relied on a strategy first put forward in \cite{Giryavets:2003vd} and exploited the discrete symmetry enhancement of certain Calabi--Yau fourfolds along special loci in the interior of the moduli space.
These loci are at finite distance with respect to the moduli space metric and correspond to smooth fourfolds.
They are, hence, qualitatively different from the singular loci at which a possible violation of the tadpole conjecture was observed in \cite{Bena:2021wyr}.
In particular, they are not expected to give rise to any additional gauge or other massless degrees of freedom.

On the symmetric locus, the F-term equations can be solved by turning on flux that is invariant under the discrete symmetry group. Phrased differently, it is possible to consistently set a large number of non-invariant flux quanta to zero while still solving the supersymmetry conditions.
Still, it is important whether such a flux choice also generates obstructions to many of the deformations orthogonal to the symmetric locus. By scanning over a large class of Calabi--Yau fourfolds we found that an invariant flux with $W_0\neq 0$ in fact leads to $n_{\rm stab}$ that is (at least) proportional to the number of non-invariant deformations, cf.~\eqref{bound}.
On the other hand, there is no reason for the invariant flux to generate a large M2-brane tadpole, as can be seen, for example, via mirror symmetry.
Hence, one does not expect the strong version of the tadpole conjecture to hold at the symmetric loci.
We illustrated this in an explicit example where it is possible to find consistent fluxes with $n_{\rm stab}/N_{\rm flux}\sim \mathcal{O}(100)$, which violates the strong form of the tadpole conjecture.
However, even though we give a strict lower bound on the number of stabilized moduli, we cannot conclusively decide if actually all of them can be  stabilized.
Still, we provide some arguments for why we believe that also the weak version of the tadpole conjecture might be evaded for $W_0\neq 0$ vacua along the symmetric locus.
We intend to revisit this issue in future work. 

We can thus summarize the state of affairs in the following way: Whereas in asymptotic regions in the vicinity of points with unipotent monodromy the linear scaling predicted of the tadpole conjecture has been confirmed \cite{Grana:2022dfw}, at other special points in field space, namely the symmetric loci considered in this note, the bound predicted by the tadpole conjecture can be violated. This is consistent with the analysis of $K3\times K3$ in \cite{Bena:2021wyr} where it was shown that the tadpole conjecture bound can be violated at gauge enhancement points in the moduli space of $K3\times K3$. The symmetric loci considered in this work can be viewed as the analogue of these gauge enhancement point for compactifications on strict $SU(4)$ holonomy manifolds. The difference is that the enhanced symmetry in the latter case is discrete and the corresponding fourfold is smooth such that there are no additional massless degrees of freedom. 

The results of this paper show that extrapolating results from the asymptotic regimes to the interior of the moduli space may not be justified in general. Still, one might wonder whether the strong version of the tadpole conjecture holds at generic points in the interior of the moduli space or whether it is only possible to evade it at special loci such as the ones discussed here.
In fact, to find consistent vacua along the symmetric locus, it was crucial that we can consistently set to zero a large number of fluxes which allowed us to find vacua with relatively small $N_{\rm flux}$.
It is an open question whether this is still possible when moving away from the symmetric locus or whether in absence of any symmetry generically a large number of integer flux quanta must be  non-vanishing.

In a similar spirit, it seems conceivable that in type IIB/M-theory compactifications there is a fundamental obstruction to finding solutions that satisfy the tadpole cancellation condition while at the same time keeping a perturbative expansion under control. This would for instance explain why in the vicinity of asymptotic regimes in moduli space it is very difficult to stabilize a large number of moduli using just the leading order periods, since the asymptotic expansion of the periods can be viewed as a perturbative expansion via mirror symmetry. On the other hand, the complex structure sector is classically exact such that we do not need to rely on such a perturbative expansions.

The situation is different once we also take the K\"ahler moduli sector into account,
which is, of course, inevitable if we aim to find realistic, stable vacua.%
\footnote{K\"ahler moduli stabilization can be avoided for special ``non-geometric'' world-sheet constructions with $h^{1,1}=0$,  as discussed in \cite{Becker:2006ks, Becker:2007dn, Bardzell:2022jfh,Becker:2022hse}.}
When including K\"ahler moduli, we \emph{have} to rely on a perturbative expansion in $\alpha'$ since in a genuine four-dimensional $\cN=1$ theory (or equivalently $\cN=2$ in three dimensions) as of now we do not have a way to do computations that are exact in $\alpha'$.
Let us stress that in this work we exclusively focused on the complex structure sector and secretly assumed that we can treat this sector completely decoupled from the K\"ahler sector.
However, in a genuine $\cN=1$ theory the K\"ahler and complex structure sectors do not decouple at finite volume (see, e.g., \cite{Wiesner:2022qys} for a recent discussion in F-theory).
Therefore, any discussion of complex structure moduli stabilization is strictly speaking only valid in the large volume or decompactification limit, unless amended by an analysis of the K\"ahler moduli sector. Since such an analysis necessarily relies on a perturbative $\alpha'$-expansion, we would in general expect a tension between control over this perturbative expansion and the tadpole cancellation condition. It might not come as a surprise that constraints arising from the tadpole cancellation are intimately related to the validity of a perturbative expansion, most prominently the $\alpha'$-expansion. After all, in M-theory the tadpole cancellation condition is derived from a higher-derivative term and is therefore part of the $\alpha'$-expansion. The tension between tadpole cancellation and perturbative control is essentially what was identified as a possible obstruction to the type IIB LVS scenario \cite{Balasubramanian:2005zx} in \cite{Junghans:2022exo,Gao:2022fdi, Junghans:2022kxg}. Similarly, in the context of the KKLT scenario \cite{Kachru:2003aw}, a tension between tadpole cancellation and the existence of perturbatively controlled supersymmetric AdS flux vacua and potential issues with a controlled antibrane uplift  were pointed out in \cite{Bena:2018fqc,Blumenhagen:2019qcg,Carta:2019rhx,Bena:2019sxm,Randall:2019ent,Dudas:2019pls,Gao:2020xqh,Lust:2022lfc,Lust:2022xoq,Blumenhagen:2022dbo}.

Flux compactifications with $W_0 \neq 0$ play a particularly prominent role in the construction of vacua \`a la KKLT. 
There, the non-vanishing, classical contribution $W_0$ to the superpotential gets balanced against additional, non-perturbative terms in order to achieve also K\"ahler moduli stablization.
In this context it is crucial that $W_0$ takes an exponentially small value such that control over the non-perturbative correction terms can be maintained (see \cite{Demirtas:2019sip} for a recent construction of flux vacua with $\left|W_0\right| \ll 1$).
On the other hand, in our setup we have seen that there is generally a large number of complex structure moduli that obtain classical masses of order $m \sim \left|W_0\right|$.
To be more specific, this is the case for all non-invariant moduli that lie in the kernel of $D_i D_j W$.
However, according to figure~\ref{fig:rankM1} the dimension of $\ker \left(D_i D_j W\right)$ is directly proportional to the number of moduli, with an order-one proportionality factor.
Therefore, there is generally a large number of complex structure moduli with masses comparable to those of the K\"ahler moduli and the size of the non-perturbative corrections.
It is hence not clear if it is possible to integrate these moduli out and to discuss K\"ahler moduli and complex structure moduli stabilization independently, as originally proposed in \cite{Kachru:2003aw} (for a related discussion of KKLT in the presence of very light complex structure moduli see, e.g., \cite{Blumenhagen:2019qcg,Dudas:2019pls,Blumenhagen:2022dbo}).

In general it would be interesting to see whether the symmetric loci discussed in this work actually do survive as part of the moduli space even away from large volume or whether the mixture between K\"ahler and complex structure sector becomes important in the vicinity of the symmetric loci.

\subsubsection*{Acknowledgements}
We would like to thank Mikel Alvarez, Iosif Bena, Thibaut Coudarchet, Naomi Gendler, Mariana Gra\~na, Thomas Grimm, Damian van de Heisteeg, Alvaro Herr\'aez, Fernando Marchesano, David Prieto, and Cumrun Vafa for useful discussions and correspondence. 
In addition we would like to thank the Simons Center for Geometry and Physics, Stony Brook University for hospitality at the 2022 summer workshop where parts of this work were done.
The work of SL is supported by the NSF grant PHY-1915071.
The work of MW is supported in part by a
grant from the Simons Foundation (602883, CV) and also by the NSF grant PHY-2013858.

%
%
%
%
%
%
%
%
\bibliography{papers_Max}

\providecommand{\href}[2]{#2}\begingroup\raggedright\begin{thebibliography}{10}

\bibitem{Ooguri:2006in}
H.~Ooguri and C.~Vafa, {\it {On the Geometry of the String Landscape and the
  Swampland}},  {\em Nucl. Phys.} {\bf B766} (2007) 21--33,
  [\href{http://arxiv.org/abs/hep-th/0605264}{{\tt hep-th/0605264}}].

\bibitem{Brennan:2017rbf}
T.~D. Brennan, F.~Carta, and C.~Vafa, {\it {The String Landscape, the
  Swampland, and the Missing Corner}},  {\em PoS} {\bf TASI2017} (2017) 015,
  [\href{http://arxiv.org/abs/1711.00864}{{\tt arXiv:1711.00864}}].

\bibitem{Palti:2019pca}
E.~Palti, {\it {The Swampland: Introduction and Review}},  {\em Fortsch. Phys.}
  {\bf 67} (2019), no.~6 1900037, [\href{http://arxiv.org/abs/1903.06239}{{\tt
  arXiv:1903.06239}}].

\bibitem{vanBeest:2021lhn}
M.~van Beest, J.~Calder\'on-Infante, D.~Mirfendereski, and I.~Valenzuela, {\it
  {Lectures on the Swampland Program in String Compactifications}},  {\em Phys.
  Rept.} {\bf 989} (2022) 1--50, [\href{http://arxiv.org/abs/2102.01111}{{\tt
  arXiv:2102.01111}}].

\bibitem{Grana:2021zvf}
M.~Gra\~na and A.~Herr\'aez, {\it {The Swampland Conjectures: A Bridge from
  Quantum Gravity to Particle Physics}},  {\em Universe} {\bf 7} (2021), no.~8
  273, [\href{http://arxiv.org/abs/2107.00087}{{\tt arXiv:2107.00087}}].

\bibitem{Grana:2005jc}
M.~Gra\~na, {\it {Flux compactifications in string theory: A Comprehensive
  review}},  {\em Phys. Rept.} {\bf 423} (2006) 91--158,
  [\href{http://arxiv.org/abs/hep-th/0509003}{{\tt hep-th/0509003}}].

\bibitem{Douglas:2003um}
M.~R. Douglas, {\it {The Statistics of string / M theory vacua}},  {\em JHEP}
  {\bf 05} (2003) 046, [\href{http://arxiv.org/abs/hep-th/0303194}{{\tt
  hep-th/0303194}}].

\bibitem{Denef:2004ze}
F.~Denef and M.~R. Douglas, {\it {Distributions of flux vacua}},  {\em JHEP}
  {\bf 05} (2004) 072, [\href{http://arxiv.org/abs/hep-th/0404116}{{\tt
  hep-th/0404116}}].

\bibitem{Denef:2004cf}
F.~Denef and M.~R. Douglas, {\it {Distributions of nonsupersymmetric flux
  vacua}},  {\em JHEP} {\bf 03} (2005) 061,
  [\href{http://arxiv.org/abs/hep-th/0411183}{{\tt hep-th/0411183}}].

\bibitem{Bena:2020xrh}
I.~Bena, J.~Bl\r{a}b\"ack, M.~Gra\~na, and S.~L\"ust, {\it {The tadpole
  problem}},  {\em JHEP} {\bf 11} (2021) 223,
  [\href{http://arxiv.org/abs/2010.10519}{{\tt arXiv:2010.10519}}].

\bibitem{Bena:2021wyr}
I.~Bena, J.~Bl\r{a}b\"ack, M.~Gra\~na, and S.~L\"ust, {\it {Algorithmically
  Solving the Tadpole Problem}},  {\em Adv. Appl. Clifford Algebras} {\bf 32}
  (2022), no.~1 7, [\href{http://arxiv.org/abs/2103.03250}{{\tt
  arXiv:2103.03250}}].

\bibitem{Marchesano:2021gyv}
F.~Marchesano, D.~Prieto, and M.~Wiesner, {\it {F-theory flux vacua at large
  complex structure}},  {\em JHEP} {\bf 08} (2021) 077,
  [\href{http://arxiv.org/abs/2105.09326}{{\tt arXiv:2105.09326}}].

\bibitem{Plauschinn:2021hkp}
E.~Plauschinn, {\it {The tadpole conjecture at large complex-structure}},  {\em
  JHEP} {\bf 02} (2022) 206, [\href{http://arxiv.org/abs/2109.00029}{{\tt
  arXiv:2109.00029}}].

\bibitem{Lust:2021xds}
S.~L\"ust, {\it {Large complex structure flux vacua of IIB and the Tadpole
  Conjecture}},  \href{http://arxiv.org/abs/2109.05033}{{\tt
  arXiv:2109.05033}}.

\bibitem{Grimm:2021ckh}
T.~W. Grimm, E.~Plauschinn, and D.~van~de Heisteeg, {\it {Moduli stabilization
  in asymptotic flux compactifications}},  {\em JHEP} {\bf 03} (2022) 117,
  [\href{http://arxiv.org/abs/2110.05511}{{\tt arXiv:2110.05511}}].

\bibitem{Grana:2022dfw}
M.~Gra\~na, T.~W. Grimm, D.~van~de Heisteeg, A.~Herraez, and E.~Plauschinn,
  {\it {The tadpole conjecture in asymptotic limits}},  {\em JHEP} {\bf 08}
  (2022) 237, [\href{http://arxiv.org/abs/2204.05331}{{\tt arXiv:2204.05331}}].

\bibitem{Giryavets:2003vd}
A.~Giryavets, S.~Kachru, P.~K. Tripathy, and S.~P. Trivedi, {\it {Flux
  compactifications on Calabi-Yau threefolds}},  {\em JHEP} {\bf 04} (2004)
  003, [\href{http://arxiv.org/abs/hep-th/0312104}{{\tt hep-th/0312104}}].

\bibitem{Denef:2004dm}
F.~Denef, M.~R. Douglas, and B.~Florea, {\it {Building a better racetrack}},
  {\em JHEP} {\bf 06} (2004) 034,
  [\href{http://arxiv.org/abs/hep-th/0404257}{{\tt hep-th/0404257}}].

\bibitem{Cicoli:2013cha}
M.~Cicoli, D.~Klevers, S.~Krippendorf, C.~Mayrhofer, F.~Quevedo, and
  R.~Valandro, {\it {Explicit de Sitter Flux Vacua for Global String Models
  with Chiral Matter}},  {\em JHEP} {\bf 05} (2014) 001,
  [\href{http://arxiv.org/abs/1312.0014}{{\tt arXiv:1312.0014}}].

\bibitem{Demirtas:2019sip}
M.~Demirtas, M.~Kim, L.~Mcallister, and J.~Moritz, {\it {Vacua with Small Flux
  Superpotential}},  {\em Phys. Rev. Lett.} {\bf 124} (2020), no.~21 211603,
  [\href{http://arxiv.org/abs/1912.10047}{{\tt arXiv:1912.10047}}].

\bibitem{Blanco-Pillado:2020wjn}
J.~J. Blanco-Pillado, K.~Sousa, M.~A. Urkiola, and J.~M. Wachter, {\it {Towards
  a complete mass spectrum of type-IIB flux vacua at large complex structure}},
   {\em JHEP} {\bf 04} (2021) 149, [\href{http://arxiv.org/abs/2007.10381}{{\tt
  arXiv:2007.10381}}].

\bibitem{Blanco-Pillado:2020hbw}
J.~J. Blanco-Pillado, K.~Sousa, M.~A. Urkiola, and J.~M. Wachter, {\it
  {Universal Class of Type-IIB Flux Vacua with Analytic Mass Spectrum}},  {\em
  Phys. Rev. D} {\bf 103} (2021), no.~10 106006,
  [\href{http://arxiv.org/abs/2011.13953}{{\tt arXiv:2011.13953}}].

\bibitem{Braun:2020jrx}
A.~P. Braun and R.~Valandro, {\it {$G_{4}$ flux, algebraic cycles and complex
  structure moduli stabilization}},  {\em JHEP} {\bf 01} (2021) 207,
  [\href{http://arxiv.org/abs/2009.11873}{{\tt arXiv:2009.11873}}].

\bibitem{Becker:2022hse}
K.~Becker, E.~Gonzalo, J.~Walcher, and T.~Wrase, {\it {Fluxes, Vacua, and
  Tadpoles meet Landau-Ginzburg and Fermat}},
  \href{http://arxiv.org/abs/2210.03706}{{\tt arXiv:2210.03706}}.

\bibitem{deAlwis:2013jaa}
S.~de~Alwis, J.~Louis, L.~McAllister, H.~Triendl, and A.~Westphal, {\it {Moduli
  spaces in $AdS_4$ supergravity}},  {\em JHEP} {\bf 05} (2014) 102,
  [\href{http://arxiv.org/abs/1312.5659}{{\tt arXiv:1312.5659}}].

\bibitem{Witten:1996md}
E.~Witten, {\it {On flux quantization in M theory and the effective action}},
  {\em J. Geom. Phys.} {\bf 22} (1997) 1--13,
  [\href{http://arxiv.org/abs/hep-th/9609122}{{\tt hep-th/9609122}}].

\bibitem{Braun:2014xka}
A.~P. Braun and T.~Watari, {\it {The Vertical, the Horizontal and the Rest:
  anatomy of the middle cohomology of Calabi-Yau fourfolds and F-theory
  applications}},  {\em JHEP} {\bf 01} (2015) 047,
  [\href{http://arxiv.org/abs/1408.6167}{{\tt arXiv:1408.6167}}].

\bibitem{Gukov:1999ya}
S.~Gukov, C.~Vafa, and E.~Witten, {\it {CFT's from Calabi-Yau four folds}},
  {\em Nucl. Phys. B} {\bf 584} (2000) 69--108,
  [\href{http://arxiv.org/abs/hep-th/9906070}{{\tt hep-th/9906070}}]. [Erratum:
  Nucl.Phys.B 608, 477--478 (2001)].

\bibitem{Lust:2022lfc}
S.~L\"ust, C.~Vafa, M.~Wiesner, and K.~Xu, {\it {Holography and the KKLT
  scenario}},  {\em JHEP} {\bf 10} (2022) 188,
  [\href{http://arxiv.org/abs/2204.07171}{{\tt arXiv:2204.07171}}].

\bibitem{Candelas:1989hd}
P.~Candelas, M.~Lynker, and R.~Schimmrigk, {\it {Calabi-Yau Manifolds in
  Weighted P(4)}},  {\em Nucl. Phys. B} {\bf 341} (1990) 383--402.

\bibitem{Greene:1991we}
B.~R. Greene, S.~S. Roan, and S.-T. Yau, {\it {Geometric singularities and
  spectra of Landau-Ginzburg models}},  {\em Commun. Math. Phys.} {\bf 142}
  (1991) 245--260.

\bibitem{Klemm:1996ts}
A.~Klemm, B.~Lian, S.~S. Roan, and S.-T. Yau, {\it {Calabi-Yau fourfolds for M
  theory and F theory compactifications}},  {\em Nucl. Phys. B} {\bf 518}
  (1998) 515--574, [\href{http://arxiv.org/abs/hep-th/9701023}{{\tt
  hep-th/9701023}}].

\bibitem{Vafa:1989ih}
C.~Vafa, {\it {Quantum Symmetries of String Vacua}},  {\em Mod. Phys. Lett. A}
  {\bf 4} (1989) 1615.

\bibitem{Vafa:1989xc}
C.~Vafa, {\it {String Vacua and Orbifoldized L-G Models}},  {\em Mod. Phys.
  Lett. A} {\bf 4} (1989) 1169.

\bibitem{Greene:1990ud}
B.~R. Greene and M.~R. Plesser, {\it {Duality in {Calabi-Yau} Moduli Space}},
  {\em Nucl. Phys. B} {\bf 338} (1990) 15--37.

\bibitem{Candelas:1990rm}
P.~Candelas, X.~C. De~La~Ossa, P.~S. Green, and L.~Parkes, {\it {A Pair of
  Calabi-Yau manifolds as an exactly soluble superconformal theory}},  {\em
  Nucl. Phys. B} {\bf 359} (1991) 21--74.

\bibitem{Candelas:1993dm}
P.~Candelas, X.~De~La~Ossa, A.~Font, S.~H. Katz, and D.~R. Morrison, {\it
  {Mirror symmetry for two parameter models. 1.}},  {\em Nucl. Phys. B} {\bf
  416} (1994) 481--538, [\href{http://arxiv.org/abs/hep-th/9308083}{{\tt
  hep-th/9308083}}].

\bibitem{FermatCY}
S.~L\"ust and M.~Wiesner, ``{FermatCY.nb}.''
  \url{https://github.com/sluest/FermatCY}, 2022.

\bibitem{Cota:2017aal}
C.~F. Cota, A.~Klemm, and T.~Schimannek, {\it {Modular Amplitudes and
  Flux-Superpotentials on elliptic Calabi-Yau fourfolds}},  {\em JHEP} {\bf 01}
  (2018) 086, [\href{http://arxiv.org/abs/1709.02820}{{\tt arXiv:1709.02820}}].

\bibitem{Tsagkaris:2022apo}
K.~Tsagkaris and E.~Plauschinn, {\it {Moduli stabilization in type IIB
  orientifolds at $h^{2,1}=50$}},  \href{http://arxiv.org/abs/2207.13721}{{\tt
  arXiv:2207.13721}}.

\bibitem{Becker:2006ks}
K.~Becker, M.~Becker, C.~Vafa, and J.~Walcher, {\it {Moduli Stabilization in
  Non-Geometric Backgrounds}},  {\em Nucl. Phys. B} {\bf 770} (2007) 1--46,
  [\href{http://arxiv.org/abs/hep-th/0611001}{{\tt hep-th/0611001}}].

\bibitem{Becker:2007dn}
K.~Becker, M.~Becker, and J.~Walcher, {\it {Runaway in the Landscape}},  {\em
  Phys. Rev. D} {\bf 76} (2007) 106002,
  [\href{http://arxiv.org/abs/0706.0514}{{\tt arXiv:0706.0514}}].

\bibitem{Bardzell:2022jfh}
J.~Bardzell, E.~Gonzalo, M.~Rajaguru, D.~Smith, and T.~Wrase, {\it {Type IIB
  flux compactifications with h$^{1,1}$ = 0}},  {\em JHEP} {\bf 06} (2022) 166,
  [\href{http://arxiv.org/abs/2203.15818}{{\tt arXiv:2203.15818}}].

\bibitem{Wiesner:2022qys}
M.~Wiesner, {\it {Light Strings and Strong Coupling in F-theory}},
  \href{http://arxiv.org/abs/2210.14238}{{\tt arXiv:2210.14238}}.

\bibitem{Balasubramanian:2005zx}
V.~Balasubramanian, P.~Berglund, J.~P. Conlon, and F.~Quevedo, {\it
  {Systematics of moduli stabilisation in Calabi-Yau flux compactifications}},
  {\em JHEP} {\bf 03} (2005) 007,
  [\href{http://arxiv.org/abs/hep-th/0502058}{{\tt hep-th/0502058}}].

\bibitem{Junghans:2022exo}
D.~Junghans, {\it {LVS de Sitter Vacua are probably in the Swampland}},
  \href{http://arxiv.org/abs/2201.03572}{{\tt arXiv:2201.03572}}.

\bibitem{Gao:2022fdi}
X.~Gao, A.~Hebecker, S.~Schreyer, and G.~Venken, {\it {The LVS parametric
  tadpole constraint}},  {\em JHEP} {\bf 07} (2022) 056,
  [\href{http://arxiv.org/abs/2202.04087}{{\tt arXiv:2202.04087}}].

\bibitem{Junghans:2022kxg}
D.~Junghans, {\it {Topological constraints in the LARGE-volume scenario}},
  {\em JHEP} {\bf 08} (2022) 226, [\href{http://arxiv.org/abs/2205.02856}{{\tt
  arXiv:2205.02856}}].

\bibitem{Kachru:2003aw}
S.~Kachru, R.~Kallosh, A.~D. Linde, and S.~P. Trivedi, {\it {De Sitter vacua in
  string theory}},  {\em Phys. Rev. D} {\bf 68} (2003) 046005,
  [\href{http://arxiv.org/abs/hep-th/0301240}{{\tt hep-th/0301240}}].

\bibitem{Bena:2018fqc}
I.~Bena, E.~Dudas, M.~Gra\~na, and S.~L\"ust, {\it {Uplifting Runaways}},  {\em
  Fortsch. Phys.} {\bf 67} (2019), no.~1-2 1800100,
  [\href{http://arxiv.org/abs/1809.06861}{{\tt arXiv:1809.06861}}].

\bibitem{Blumenhagen:2019qcg}
R.~Blumenhagen, D.~Klaewer, and L.~Schlechter, {\it {Swampland Variations on a
  Theme by KKLT}},  {\em JHEP} {\bf 05} (2019) 152,
  [\href{http://arxiv.org/abs/1902.07724}{{\tt arXiv:1902.07724}}].

\bibitem{Carta:2019rhx}
F.~Carta, J.~Moritz, and A.~Westphal, {\it {Gaugino condensation and small
  uplifts in KKLT}},  {\em JHEP} {\bf 08} (2019) 141,
  [\href{http://arxiv.org/abs/1902.01412}{{\tt arXiv:1902.01412}}].

\bibitem{Bena:2019sxm}
I.~Bena, A.~Buchel, and S.~L\"ust, {\it {Throat destabilization (for profit and
  for fun)}},  \href{http://arxiv.org/abs/1910.08094}{{\tt arXiv:1910.08094}}.

\bibitem{Randall:2019ent}
L.~Randall, {\it {The Boundaries of KKLT}},  {\em Fortsch. Phys.} {\bf 68}
  (2020), no.~3-4 1900105, [\href{http://arxiv.org/abs/1912.06693}{{\tt
  arXiv:1912.06693}}].

\bibitem{Dudas:2019pls}
E.~Dudas and S.~L\"ust, {\it {An update on moduli stabilization with antibrane
  uplift}},  {\em JHEP} {\bf 03} (2021) 107,
  [\href{http://arxiv.org/abs/1912.09948}{{\tt arXiv:1912.09948}}].

\bibitem{Gao:2020xqh}
X.~Gao, A.~Hebecker, and D.~Junghans, {\it {Control issues of KKLT}},  {\em
  Fortsch. Phys.} {\bf 68} (2020) 2000089,
  [\href{http://arxiv.org/abs/2009.03914}{{\tt arXiv:2009.03914}}].

\bibitem{Lust:2022xoq}
S.~L\"ust and L.~Randall, {\it {Effective Theory of Warped Compactifications
  and the Implications for KKLT}},  {\em Fortsch. Phys.} {\bf 70} (2022),
  no.~7-8 2200103, [\href{http://arxiv.org/abs/2206.04708}{{\tt
  arXiv:2206.04708}}].

\bibitem{Blumenhagen:2022dbo}
R.~Blumenhagen, A.~Gligovic, and S.~Kaddachi, {\it {Mass Hierarchies and
  Quantum Gravity Constraints in DKMM-refined KKLT}},
  \href{http://arxiv.org/abs/2206.08400}{{\tt arXiv:2206.08400}}.

\end{thebibliography}\endgroup
\bibliographystyle{JHEP}

\end{document}